\documentclass[twocolumn]{aastex631}

\usepackage{tabularx}
\expandafter\let\csname singlespace\endcsname\relax
\expandafter\let\csname doublespace\endcsname\relax
\usepackage{amsmath}

\usepackage{booktabs}
\usepackage{threeparttable}
\usepackage{setspace}
\usepackage{indentfirst}
\usepackage{lipsum}
\usepackage{makecell}
\usepackage{graphics} 
\usepackage{color}
\usepackage{ulem}
\usepackage{mathrsfs}
\usepackage[figuresright]{rotating}
\usepackage{ragged2e}
\usepackage{physics}
\begin{document}

\title{Constraining the PeV gamma-ray emission zone of Cygnus X-3 with contemporaneous GeV timing and spectral observations}

\author[0000-0001-6728-902X]{Xing-Fu Zhang}
\affiliation{School of Astronomy and Space Science, Nanjing University, Nanjing 210023, China}
\affiliation{Key laboratory of Modern Astronomy and Astrophysics (Nanjing University), Ministry of Education, Nanjing 210023, China}

\author[0000-0003-1576-0961]{Ruo-Yu Liu}
\affiliation{School of Astronomy and Space Science, Nanjing University, Nanjing 210023, China}
\affiliation{Key laboratory of Modern Astronomy and Astrophysics (Nanjing University), Ministry of Education, Nanjing 210023, China}
\affiliation{Tianfu Cosmic Ray Research Center, Chengdu 610000, Sichuan, China}

\author[0000-0002-7576-7869]{Dmitriy Khangulyan}
\affiliation{Tianfu Cosmic Ray Research Center, Chengdu 610000, Sichuan, China}
\affiliation{State Key Laboratory of Particle Astrophysics Experimental Physics Division Computing Center, Institute of High Energy Physics, Chinese Academy of Sciences,Beijing,100049,China}

\author[0000-0002-0170-0741]{Cui-Yuan Dai}
\affiliation{School of Astronomy and Space Science, Nanjing University, Nanjing 210023, China}
\affiliation{Key laboratory of Modern Astronomy and Astrophysics (Nanjing University), Ministry of Education, Nanjing 210023, China}

\author[0000-0002-5881-335X]{Xiang-Yu Wang}
\affiliation{School of Astronomy and Space Science, Nanjing University, Nanjing 210023, China}
\affiliation{Key laboratory of Modern Astronomy and Astrophysics (Nanjing University), Ministry of Education, Nanjing 210023, China}

\correspondingauthor{Ruo-Yu Liu; Dmitriy Khangulyan}
\email{ryliu@nju.edu.cn; khangulyan@ihep.ac.cn}

\begin{abstract}
Cygnus X-3 has recently been established as a variable ultra-high-energy(UHE) gamma-ray source with photons detected up to 3.7~PeV. The temporal correlation between its PeV activity and GeV flares, together with the possible orbital modulation, suggests that the emission is produced within or close to the binary system. In this work, we test whether the contemporaneous GeV emission zone can also host the acceleration of the parent protons responsible for the multi-PeV photons. We jointly model the contemporaneous \textit{Fermi}-LAT spectrum and orbital light curve with a one-zone leptonic scenario dominated by anisotropic external inverse-Compton scattering. The fit places the GeV emission region at $H\sim2.8\times10^{11}\,$cm and constrains the magnetic field--size product to $BH\lesssim10^{13.3}\,$G\,cm at the $3\sigma$ level. This implies a maximum proton energy of only $\sim0.3$~PeV from the Hillas criterion, far below that required by the observed PeV emission. We therefore conclude that the GeV zone cannot be the main PeV acceleration site. Instead, the PeV emission should originate from a more compact inner region, and the jet magnetic field must dissipate rapidly between the PeV and GeV emitting zones.
\end{abstract}

\section{Introduction} \label{sec:intro}
The origin of Galactic PeV cosmic rays (CRs), which form the so-called ``knee'' of the CR spectrum \citep{1964ocrbookG,1959On}, remains a central problem in high-energy astrophysics. While supernova remnants have long been considered the primary candidates, recent TeV--PeV gamma-ray observations increasingly point to a broader population of Galactic ``PeVatrons'', such as young massive stellar clusters \citep{2024SciBu..69..449L} and microquasars \citep{LHAASO2025_microquasar,Alfaro2024}. 

The Large High Altitude Air Shower Observatory (LHAASO) has recently reported compelling evidence that the microquasar Cygnus~X-3 is a variable UHE gamma-ray emitter, with a spectrum extending from $\sim 0.06$ to 3.7~PeV and month-scale variability \citep{2025arXivLHAASO}. This establishes Cygnus X-3 as an extreme particle accelerator and implies parent proton energies of $\gtrsim 30$~PeV if the highest-energy photons are of hadronic origin. Moreover, the UHE signal shows a $3.2\sigma$ indication of orbital modulation and is temporally correlated with GeV activity observed by \textit{Fermi}-LAT, strongly suggesting that the PeV photons are produced within, or in close proximity to, the binary system.

Cygnus~X-3 is a well-studied high-mass X-ray binary (HMXB) consisting of a compact object (likely a black hole; see \citealt{2022ApJ...926..123A}) orbiting a Wolf--Rayet (WR) star with a short 4.8 hr period \citep{1967ApJ...148L.119G,2008MNRAS.384..278H}, located at a distance of $\sim 7$--$9$\,kpc \citep{2009ApJ...695.1111L,2023ApJMiller}. It exhibits dramatic radio flares \citep{2004ApJMiller,2009Miller,2022MNRASSpencer}, launching jets aligned relatively close to the line of sight, as suggested by radio observations \citep{2024NatAsVeledina,2009Miller,2001ApJ...553..766M} and velocities spanning from sub-relativistic \citep{1996AJ....112.2690W} to mildly relativistic \citep{2004ApJMiller,2001ApJ...553..766M}. The GeV emission of Cygnus~X-3 is not steady and clearly modulated at the orbital period \citep{2009SciFermi,2009Natur.462..620T}. It is widely interpreted as anisotropic external inverse-Compton (EC) scattering of the companion's intense photon field by non-thermal electrons in the jet \citep{2010MNRASDubus}. 

Modeling of the GeV orbital modulation has provided constraints on the GeV emission site and its magnetic field. Previous studies \citep{2012MNRASZdziarski,2018MNRASZdziarski,2024ApJDmytriiev} typically locate the EC region at a few orbital radii from the compact object and require a comparatively low magnetic field ($\lesssim 10$--$100\,$G), although the best-fit results depend on the detailed shape of the GeV periodic light curve, which may vary between different activity episodes. These constraints, when confronted with the new LHAASO discovery, raise a potential difficulty for accelerating protons to the required energies. This can be illustrated by the Hillas criterion \citep{Hillas1984},
\begin{equation}
E_\text{p,max}=15\left(\frac{B}{10^3~\mathrm{G}}\right)\left(\frac{H}{10^{12}~\mathrm{cm}}\right)
\left(\frac{\alpha}{0.1}\right)\left(\frac{\beta_\mathrm{j}}{0.5}\right) \,\rm PeV,
\label{eq:hillas}
\end{equation}
where $B$ is the magnetic field strength and $H$ is the distance of the acceleration zone from the compact object, while $\alpha$ is the jet opening angle and $\beta_\mathrm{j}$ is the bulk velocity in units of the speed of light.

In this work, we focus on the \textit{Fermi}-LAT data contemporaneous with the LHAASO PeV high state to determine whether the GeV emission region can simultaneously serve as the main PeV acceleration site. Specifically, we perform a joint fit to the phase-averaged GeV spectrum and the orbital light curve using an anisotropic EC model, thereby constraining the jet geometry, emission location, and magnetic field during the relevant epoch. We then map the profile-\(\chi^2\) landscape in the \((B,H)\) plane and use it to test whether the parameter region capable of satisfying the Hillas requirement for PeV proton acceleration is compatible with the contemporaneous GeV observations.

\section{Model Description}
\label{sec:model}

\subsection{Geometry and Coordinate System}
\begin{figure}[htbp]
    \centering
    \includegraphics[width=0.5\textwidth]{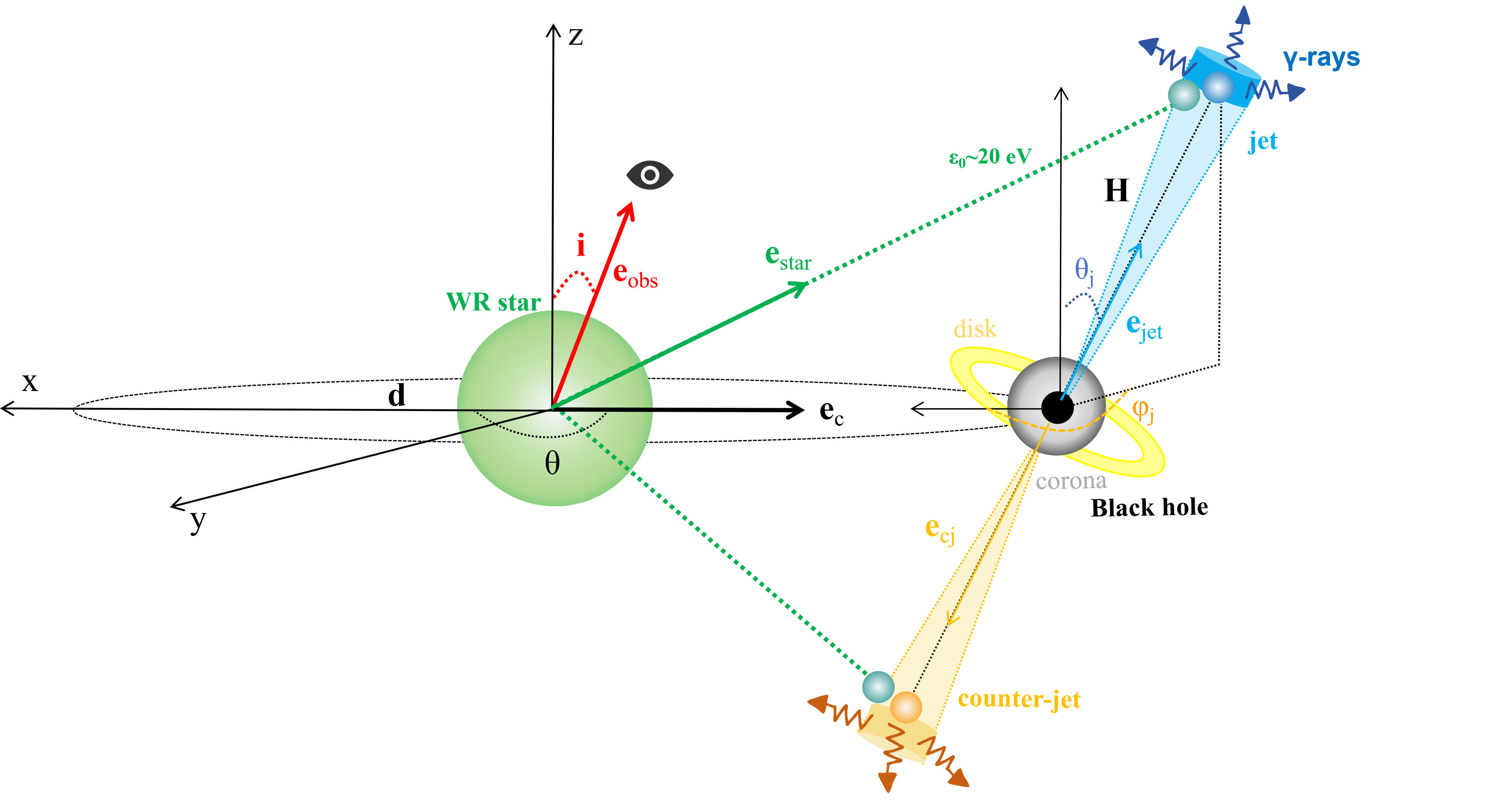}
    \caption{Schematic representation of the Cygnus X-3 geometry and coordinate system (not to scale).}
    \label{fig:geometry}
\end{figure}

The geometric framework of the system is illustrated in Figure~\ref{fig:geometry}. We adopt a jet-launching geometry based on established models for binary systems \citep{2010MNRASDubus,2012MNRASZdziarski,2018MNRASZdziarski}. The observer views the binary system at an inclination $i=30^{\circ}$ \citep{2022ApJ...926..123A}. The direction of the observer's line of sight is denoted by the unit vector $\boldsymbol{e}_{\mathrm{obs}}=(-\sin i, 0, \cos i)$. The time-dependent orientation of the system is parameterized by the orbital phase $\phi \in [0,1]$. We define the coordinate system such that the soft X-ray flux minimum (orbital phase $\phi=0$) corresponds to the superior conjunction of the compact object, where it is located behind the donor star from the observer's point of view \citep{2025arXivLHAASO}. Assuming a circular orbit, the true-anomaly angle used in our geometric calculations is then $\theta=2\pi\phi$. Consequently, the unit vector pointing from the donor star to the compact object is given by $\boldsymbol{e}_{\mathrm{c}}=(\cos\theta, \sin\theta, 0)$.

The jet orientation is defined by the polar angle $\theta_{\mathrm{j}}$ ($0\le\theta_{\mathrm{j}}\le\pi/2$) relative to the binary axis and the azimuthal angle $\phi_{\mathrm{j}}$ ($0\le\phi_{\mathrm{j}}\le2\pi$) projected onto the orbital plane. The jet direction vector is $\boldsymbol{e}_{\mathrm{j}}=(\cos\phi_{\mathrm{j}} \sin\theta_{\mathrm{j}}, \sin\phi_{\mathrm{j}} \sin\theta_{\mathrm{j}}, \cos\theta_{\mathrm{j}})$. We also account for the counter-jet geometry, defined by $\boldsymbol{e}_{\mathrm{cj}}=-\boldsymbol{e}_{\mathrm{j}}$. The GeV emission region is located at a distance $H$ from the compact object along the jet axis. The unit vector pointing from the donor star to the emission region is $\boldsymbol{e}_*=(a\boldsymbol{e}_{\mathrm{c}}+H\boldsymbol{e}_{\mathrm{j}})/R$, where $R=\qty[a^2+H^2+2aH(\boldsymbol{e}_{\mathrm{c}}\cdot\boldsymbol{e}_{\mathrm{j}})]^{1/2}$ is the distance between the star and the emission site, and $a$ is the orbital separation, taken to be $2.7\times10^{11}$~cm. The viewing angle of the jet is given by $i_{\text{jet}} = \arccos(\cos i \cos \theta_{\mathrm{j}}-\sin i\cos \phi_{\mathrm{j}}\sin \theta_{\mathrm{j}})$.

We adopt a jet semi-opening angle of $5^{\circ}$ ($\alpha \approx 0.1$), consistent with constraints from radio observations \citep{2006MNRAS.367.1432M,2022MNRASSpencer}. Furthermore, we assume that the jet orientation remains fixed over the orbital cycle.

\subsection{Radiative Model and Fitting Strategy}
We model the gamma-ray emission employing a conventional one-zone leptonic model. While we adopt the formalism for anisotropic EC from \citet{2012MNRASZdziarski}, we follow a different description of the non-thermal electron population: instead of assuming a fixed steady-state spectrum, we assume a continuous, phase-independent power-law injection of electrons:
\begin{equation}
Q(\gamma) = K_{\text{inj}} \gamma^{-p} \quad \text{for } \gamma_{1} \le \gamma \le \gamma_{2},
\end{equation}
where $K_{\text{inj}}$ is the constant normalization factor, $p$ is the power-law index, and $[\gamma_{1},\gamma_{2}]$ defines the energy range, with $\gamma_{2}$ fixed at $10^{6}$.

The resulting steady-state electron spectrum, $N(\gamma, \theta)$, varies with $\theta$. This phase dependence arises because the electron cooling rate is dominated by EC scattering of stellar photons and is therefore phase dependent. Both the distance between the emission zone and the star, $R$, and the Doppler factor of the emission zone relative to the star are sensitive to orbital phase. We compute the steady-state electron distribution by balancing continuous injection against radiative losses, including synchrotron radiation, synchrotron self-Compton (SSC), EC, and adiabatic expansion. The companion star is treated as a blackbody emitter with $T_*=10^5$~K and a photospheric radius of $R_*=10^{11}$~cm \citep{2007ARA&A..45..177C}. The system is located at a distance of $D=9~\mathrm{kpc}$ from Earth \citep{2023ApJMiller}.

To constrain the GeV emission region, we jointly fit the phase-averaged $\gamma$-ray spectrum and the orbital light curve. We perform a grid scan in the $(B,H)$ plane, and at each grid point optimize the remaining model parameters by minimizing the total $\chi^2$ of the spectrum and light curve, thereby obtaining a profile-$\chi^2$ map, $\chi^2_{\min}(B,H)$. The electron injection spectral index is allowed to vary within $p \in [3,5]$. This range is motivated by the observed GeV spectral slope: the contemporaneous \textit{Fermi}-LAT spectrum implies $p\sim 3.4$ in the fast-cooling regime and $p\sim 4.4$ in the slow-cooling regime for IC emission in the Thomson regime. This range is also consistent with previous modeling of Cygnus X-3 \citep{2010MNRASDubus,2024ApJDmytriiev,2012MNRASZdziarski}. The normalization $K_{\text{inj}}$ is constrained such that the total electron injection power does not exceed $L_{\mathrm{e,inj}}\leq10^{39}~\mathrm{erg~s^{-1}}$, motivated by X-ray polarimetry estimates of the accretion power \citep{2024NatAsVeledina}. We further restrict the minimum electron Lorentz factor to $\gamma_1 \in [10^2,10^4]$; values below $10^2$ are disfavored because they would overproduce X-rays via IC scattering of stellar photons \citep{2012MNRASZdziarski,2011A&A...529A.120C}, while values above $10^4$ would suppress the seed electrons required for the GeV emission. This procedure is designed to reconstruct the physical conditions of the GeV emission region directly implied by the contemporaneous \textit{Fermi}-LAT spectrum and orbital modulation during the LHAASO PeV epoch. We then examine whether this observationally inferred GeV zone, rather than an arbitrarily assumed jet region, is capable of confining and accelerating protons to the energies required for the observed multi-PeV photons. We also perform an MCMC exploration to estimate parameter uncertainties. In Table~\ref{tab:parameters}, the $(B,H)$ constraint is taken from the profile-$\chi^2$ scan, while the quoted 1$\sigma$ ranges of the remaining parameters are inferred from the MCMC posterior distributions.  Any derived jet-related quantities are then calculated from these fitted parameters. 

\begin{table*}[htbp]
    \centering
    \small
    \setlength{\tabcolsep}{4pt}
    \caption{Best-fit parameters from the joint fit to the GeV spectrum and orbital light curve, together with the derived constraint on the magnetic field--size product $BH$ in the contemporaneous GeV emission region. The quoted 1$\sigma$ ranges of $B$ and $H$ are estimated from the projection of the 1$\sigma$ confidence contour in the $(B,H)$ plane, while those of the remaining parameters are inferred from the MCMC posterior distributions.}
    \label{tab:parameters}
    \begin{threeparttable}
        \begin{tabular*}{\textwidth}{@{\extracolsep{\fill}} clccc}
            \toprule
            \textbf{Parameter} & \textbf{Explanation} & \textbf{Unit} & \textbf{Range} & \textbf{Best fit / constraint} \\
            \midrule
            $H$ & Distance of the GeV emission zone along the jet & cm & $[10^{11},3\times10^{12}]$ & $(2.8^{+2.1}_{-0.8}) \times 10^{11}$ \\
            $\beta_j$ & Bulk velocity of the relativistic jet & $c$ & $[0.01,0.99]$ & $0.55^{+0.15}_{-0.15}$ \\
            $\phi_j$ & Azimuthal angle of the jet projection & $^\circ$ & $[0,360]$ & $196^{+13}_{-3}$ \\
            $\theta_j$ & Inclination angle of the jet axis & $^\circ$ & $[0,90]$ & $46^{+13}_{-11}$ \\
            $p$ & Electron spectral index & -- & $[3,5]$ & $3.6^{+0.2}_{-0.2}$ \\
            $K_{\text{inj}}$ & Electron injection normalization & s$^{-1}$ & $[10^{45},10^{50}]$ & $(1.4^{+12.0}_{-1.3})\times10^{48}$ \\
            $\gamma_{1}$ & Minimum Lorentz factor of injected electrons & -- & $[10^2,10^4]$ & $2400^{+990}_{-950}$ \\
            $B$ & Magnetic field strength in the GeV emission region & G & $[10^{-2},10^4]$ & $20^{+32.4}_{-19.9}$ \\
            $BH^*$ & Magnetic field--size product in the GeV emission region & G\,cm & -- & $\leq 10^{13.3}$ \\
            \bottomrule
        \end{tabular*}
        \begin{tablenotes}
            \item *The quoted 3$\sigma$ upper limit on $BH$ is derived from the $\chi^2$-profile scan in the $(B,H)$ plane.
        \end{tablenotes}
    \end{threeparttable}
\end{table*}

\section{Results}
\label{sec:results}
\begin{figure}[htbp]
    \centering
    \includegraphics[width=0.5\textwidth]{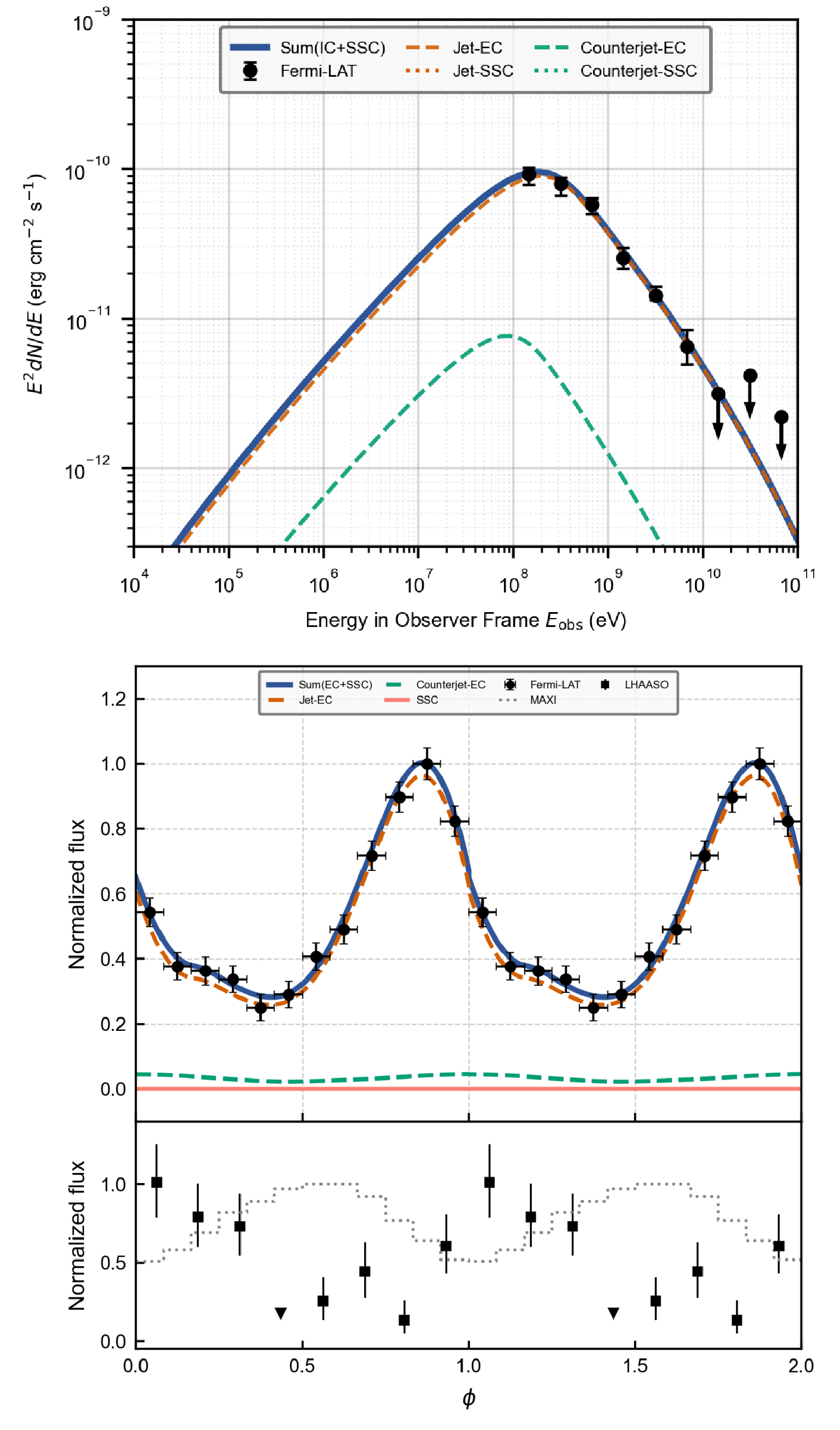}
\caption{Broadband modeling of Cygnus~X-3 during the LHAASO PeV epoch. \textbf{Upper panel:} \textit{Fermi}-LAT spectral energy distribution. Black circles and arrows denote measured fluxes and upper limits. The solid blue curve shows the total best-fit model, composed of EC emission from the jet (orange dashed) and counter-jet (green dashed). \textbf{Lower panel:} Normalized orbital light curves. Two orbital periods are shown to visualize the phase continuity. \textit{Fermi}-LAT ($0.1$--$100$~GeV; black circles) data used in the fit are shown together with the LHAASO ($\ge 0.1$~PeV; black squares) and MAXI ($2$--$20$~keV; gray dotted curve) light curves for comparison of their relative phase behavior.}
\label{fig:benchmark}
\end{figure}

Our one-zone leptonic model successfully reproduces the phase-averaged GeV spectrum and the orbital light curve (Figure~\ref{fig:benchmark}),  yielding a minimum $\chi^2$ of $\chi^2_{\min}=4$. The fit includes 6 detected spectral data points and 12 light-curve points, corresponding to 10 degrees of freedom for the global best-fit solution. The quoted $\chi^2$ values are computed using detected points only. Upper limits are used to exclude models that would clearly exceed them and for visual comparison, but they are not included in the fit statistic. The emission is dominated by anisotropic EC scattering of stellar photons from the jet and counter-jet. We also verified that a small eccentricity ($e\approx0.03$; \citealt{2019ApJ...871..244A}) does not affect the main conclusions (see Appendix~\ref{app:Eccentricity}).

The best-fit solution favors a moderately relativistic jet with $\beta_j=0.55^{+0.15}_{-0.15}$, consistent with the proper-motion estimate of \citet{2004ApJMiller}. The GeV emission region is located at $H=(2.8^{+2.1}_{-0.8})\times10^{11}$~cm, comparable to the orbital separation. The best-fit jet orientation, $(\theta_{\mathrm{j}},\phi_{\mathrm{j}})=\qty((46^{+13}_{-11})^{\circ},(196^{+13}_{-3})^{\circ})$, implies a viewing angle of $i_{\mathrm{jet}}\approx18^{\circ}$, still indicating a jet oriented relatively close to the line of sight. The electron injection is steep, with $p=3.6^{+0.2}_{-0.2}$ and $\gamma_1=2400^{+990}_{-950}$, the latter being constrained by the spectral turnover near $E_{\mathrm{break}}\sim150$~MeV. The inferred electron injection and kinetic powers are $L_{\mathrm{e,inj}}\approx3.8\times10^{36}$ and $L_{\mathrm{e,K}}\approx2.0\times10^{36}$~erg~s$^{-1}$, respectively. The corresponding cooling time at $\gamma_1$ is $\langle t_{\mathrm{cool}}\rangle\approx2.8$~s \citep[see, e.g.][for convenient expressions for IC cooling time]{2014ApJ...783..100K}, implying a compact emitting length of $h_{\mathrm{cool}}\sim4.5\times10^{10}$~cm ($\sim0.2a$), while the radial Thomson optical depth remains low, $\tau_{\mathrm{T}}\approx2.0\times10^{-5}$.

\begin{figure}[htbp]
    \centering
    \includegraphics[width=0.5\textwidth]{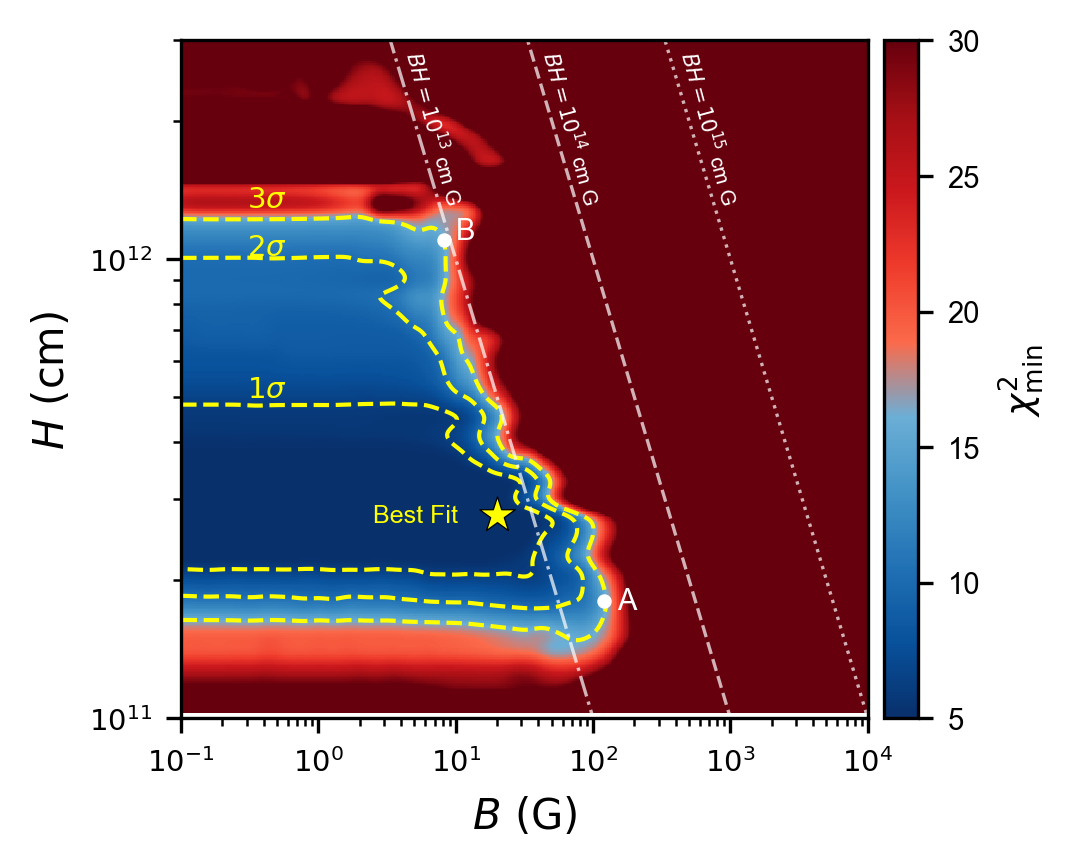}
\caption{
Profile-$\chi^2$ map in the $(B,H)$ plane. The dashed contours mark the 1$\sigma$, 2$\sigma$, and 3$\sigma$ confidence regions defined by the corresponding $\Delta\chi^2$ thresholds for two parameters of interest. The gray lines show constant $BH$. Point A denotes the largest-$BH$ solution formally enclosed by the 3$\sigma$ contour, while Point B marks the 3$\sigma$ solution that yields the global minimum decay index required to satisfy both $\tau_{\gamma B}<1$ and robust threshold $E_{\rm conf}\ge10^{16}\,\mathrm{eV}$.}
\label{fig:map}
\end{figure}

The fit favors a weakly magnetized GeV emission zone, with a best-fit magnetic field of $B\approx20$~G. This is expected because a stronger magnetic field enhances the SSC component, which is much less phase dependent than the anisotropic EC emission and therefore tends to wash out the observed orbital modulation. Figure~\ref{fig:map} shows the profile-$\chi^2$ map in the $(B,H)$ plane. The statistically allowed region is confined to relatively small values of the magnetic field-size product, with $BH\le10^{13.3}\ {\rm G\,cm}$ at the 3$\sigma$ level. Applying the Hillas criterion, even this robust upper bound implies a maximum proton energy of only $E_\text{p,max}\approx0.3~\mathrm{PeV}$ in the GeV zone, far below that required for the observed multi-PeV photons. Therefore, the GeV emission region at $H\sim a$ cannot be the main PeV proton acceleration site. For the best-fit GeV zone, the corresponding Poynting flux is $L_B\approx2.2\times10^{33}$~erg~s$^{-1}$, implying a low magnetization parameter $\sigma_{\mathrm{p}}=L_B/L_{\mathrm{e,K}}\approx1.1\times10^{-3}$. Point A in Figure~\ref{fig:map} marks the largest-$BH$ solution enclosed by the 3$\sigma$ confidence contour. Its fitted spectrum and orbital light curve are shown in Appendix~\ref{app:ABcompare}.

\section{Constraining the location of the PeV emission zone}
\label{sec:discussion}
To accelerate protons to $\sim 30$~PeV, the acceleration site must satisfy the Hillas criterion, or, more conservatively, the confinement condition that the Larmor radius of 30~PeV protons remain smaller than the jet transverse size $\alpha H$. This requirement implies a robust lower limit on the magnetic field--size product, $BH\gtrsim 10^{15}~\rm G\, cm$. Since the GeV data constrain the magnetic field and location of the downstream emission region, the magnetic-field profile along the jet becomes the key ingredient for connecting the GeV zone to the upstream PeV acceleration site. 

We parametrize the magnetic-field evolution along the jet as
\begin{equation}
B(H)=B_0\left(\frac{H_0}{H}\right)^\delta,
\label{equ:BH}
\end{equation}
taking the fitted GeV-zone parameters $(B,H)$ from Section~3 as the reference values $(B_0,H_0)$. The index $\delta$ describes the magnetic-field decay along the jet; $\delta=1$ and 2 correspond approximately to predominantly toroidal and poloidal field configurations under flux freezing. We define the characteristic proton-acceleration site as $H_1$. For $\delta=1$, the product $BH$ remains constant along the jet, so the jet never reaches the condition required for 30~PeV proton confinement. For $\delta>1$, the product $BH$ increases toward smaller radii, allowing the confinement condition to be met at some inner location $H_1<H_0$.

On the other hand, the PeV emission zone must also be transparent to magnetic pair production. The corresponding optical depth is
\begin{equation}
\tau_{\gamma B}(H)=\alpha H/\lambda(B,E_\gamma),
\label{equ:tau}
\end{equation}
where $\lambda$ is the attenuation length for a photon of energy $E_\gamma=1$~PeV in the local magnetic field. We approximate the attenuation length by smoothly connecting the high- and low-field asymptotic forms, i.e.,
$\lambda=(\lambda_1^s+\lambda_2^s)^{1/s}$ with $s=1.2$, where $\lambda_1^{-1}=1.62\times10^{-6}B_{\perp}\chi_{\gamma}^{-1/3}~\mathrm{cm}^{-1}$ for $\chi_{\gamma}\gg1$ and $\lambda_2^{-1}=9.84\times10^{-7}B_{\perp}\exp(-8/3\chi_{\gamma})~\mathrm{cm}^{-1}$ for $\chi_{\gamma}\ll1$, with $\chi_{\gamma}=E_\gamma B_{\perp}/(m_ec^2B_{\mathrm{cr}})$ \citep{WangJS2018}. Here $B_{\mathrm{cr}}=4.414\times10^{13}$~G is the critical magnetic field, and we take $B_{\perp}\approx B\sin i_{\rm jet}=0.31B$, corresponding to the case in which the magnetic field is predominantly aligned with the jet axis. 

\begin{figure}[t]
    \centering
    \includegraphics[width=0.5\textwidth]{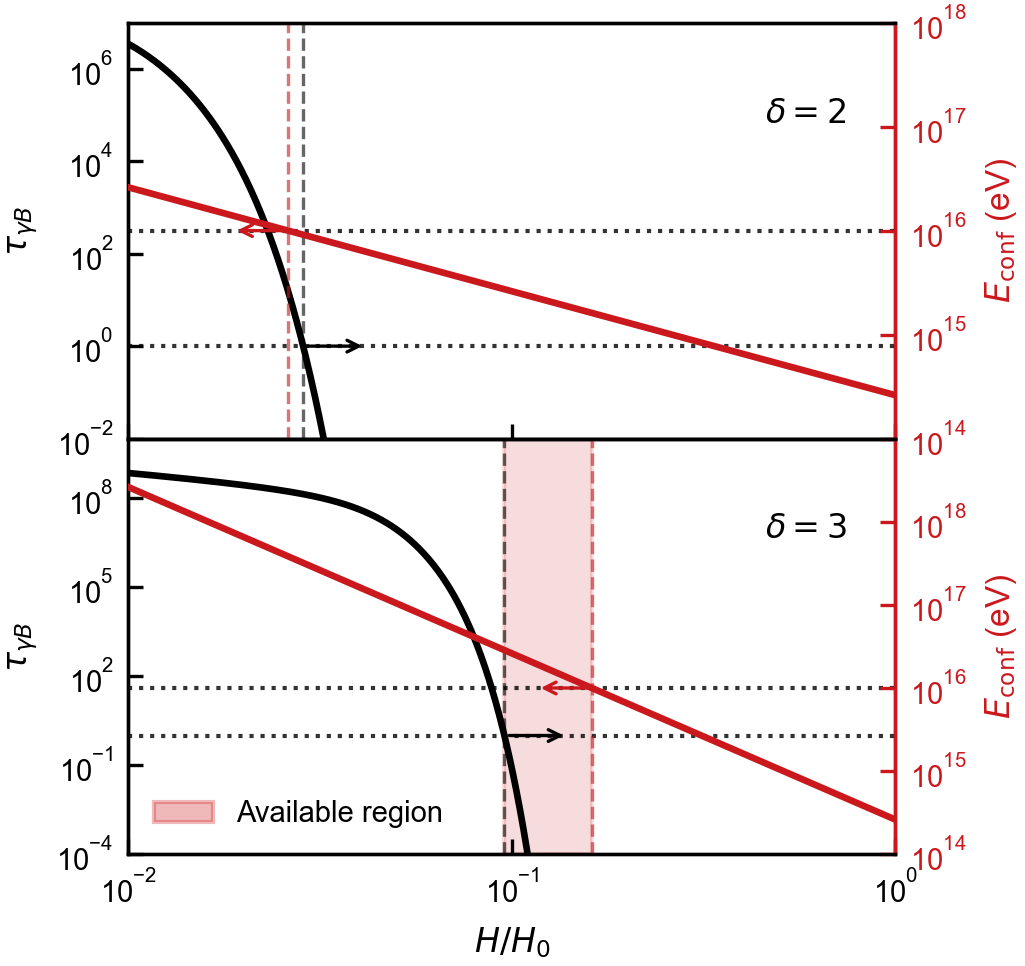}
    \caption{Constraints on the PeV emission zone for Point B in Figure~\ref{fig:map}, i.e., the 3$\sigma$ solution that yields the global minimum decay index. The two panels show the results for $\delta=2$ and 3, respectively. The black curves represent the magnetic pair-production optical depth $\tau_{\gamma B}$, and the red curves represent the confinement energy $E_{\rm conf}$. The horizontal dotted lines mark $\tau_{\gamma B}=1$ and the robust threshold $E_{\rm conf}=10^{16}\,\mathrm{eV}$. The shaded region shows where both conditions are satisfied simultaneously.}
    \label{fig:tauH}
\end{figure}

Figure~\ref{fig:map} already shows that even the 3$\sigma$ allowed region in the $(B,H)$ plane remains confined to relatively small $BH$, insufficient for 30~PeV proton confinement. The observed PeV emission is therefore likely to originate from a more compact inner jet region, where both the confinement and transparency conditions need to be satisfied. This then leads to a strong constraint on the height of the PeV emission zone. We then evaluate, for every acceptable $(B,H)$ solution within the 3$\sigma$ contour shown in Figure~\ref{fig:map}, the minimum decay index $\delta$ required to satisfy both $\tau_{\gamma B}<1$ and $E_{\rm conf}=eBH\alpha \ge10^{16}\,$eV. Here we adopt $E_{\rm conf}\ge 10^{16}\,\mathrm{eV}$ as a robust threshold for the inner-zone constraint. If even this lower threshold cannot be satisfied, then the $\sim 30~\mathrm{PeV}$ requirement is necessarily out of reach. We find that the global minimum within the 3$\sigma$ region is $\delta_{\rm min}>2$, indicating that even the most conservative solution requires a magnetic-field decay steeper than $H^{-2}$. The corresponding combination of $B$ and $H$ for this minimum scenario, $(B,H)=(8.2~\mathrm{G},\,1.1\times10^{12}~\mathrm{cm})$, is marked as Point B in Figure~\ref{fig:map}. Figure~\ref{fig:tauH} illustrates the corresponding PeV-zone constraint for this solution using $\delta=2$ and 3. For each $\delta$, we scan $H/H_0$ and evaluate both $\tau_{\gamma B}$ and $E_{\rm conf}$. The shaded region, which appears in the $\delta=3$ panel, marks where both conditions are satisfied. The allowed region lies at $H/H_0\lesssim 0.15$, implying that the PeV emission site is located at a significantly smaller distance from the compact object than the GeV emission zone. This result supports models suggesting that PeV photons are emitted close to the compact object \citep[e.g.,][]{Wei2025}.

Such an evolution is difficult to reconcile with a simple flux-freezing picture and instead points to efficient dissipation of the magnetic field between the inner jet and the orbital-scale GeV emission zone. In this interpretation, multi-PeV protons are accelerated close to the compact object, in a region of substantial magnetization, and the magnetic field then undergoes strong dissipation, potentially through magnetic reconnection, leaving behind the weakly magnetized environment inferred at $H\sim a$. Note that the actual magnetic-field profile need not follow a single power law exactly; however, the essential requirement remains the same: the field must dissipate rapidly in the inner jet in order to allow both proton confinement to $\sim30$~PeV and PeV-photon transparency.

The necessity of a compact inner PeV emission zone is also compatible with the multi-wavelength orbital light curves. By definition, the soft X-ray minimum corresponds to the superior conjunction of the compact object, where the inner accretion flow is most strongly obscured by the companion star. In contrast, the GeV--PeV $\gamma$-rays in our picture are powered by the interaction between relativistic particles and the companion radiation field, and therefore the orbital modulation is primarily governed by the anisotropic scattering geometry\footnote{The GeV--PeV emission is also dependent on the distance from the companion star to the radiation zone, which determines the density of the companion radiation. The influence of the distance, however, is subordinate for the considered parameters.}.

Importantly, once the $\gamma$-ray emission zone is displaced from the orbital plane to a finite distance $H$ along a tilted jet, the orbital phase of the maximum/minimum scattering angle (equivalently, the extrema of $\mu \equiv \cos\Psi = \boldsymbol{e}_* \cdot \boldsymbol{e}_{\mathrm{obs}}$) is no longer identical to the conjunction phase of the compact object. For the best-fit GeV geometry ($H_0 \simeq 2.8 \times 10^{11}$~cm, $\theta_{\mathrm{j}} \simeq 46^\circ$, and $\phi_{\mathrm{j}} \simeq 196^\circ$), the smallest/largest scattering angles, i.e.\ the maximum/minimum $\mu(\phi)$, occur at phases $\phi_{\min/\max} \simeq 0.41 / 0.89$, corresponding to the highest/lowest EC scattering rates. These phases are significantly offset from the inferior and superior conjunctions. As $H$ decreases, $\phi_{\min/\max}$ smoothly approach the conjunction values, recovering $\phi_{\min/\max}=0.5/0.0$ in the limit $H\to0$. This implies that a more compact emitter naturally peaks closer to the X-ray minimum phase ($\phi=0.0$). Detailed calculations in the Appendix~\ref{app:ratio} show that this geometric effect yields a phase shift of order $\sim0.1$ for $H\sim H_0$, while the shift rapidly decreases inward. Therefore, the observation that the LHAASO UHE light curve peaks close to the X-ray minimum (see the lower panel of Figure~\ref{fig:benchmark}), while the GeV peak precedes the UHE peak by a phase lag of $\Delta\phi \sim 0.08 - 0.29$, is consistent with the inferred geometry in which the PeV emission zone is located at a closer distance from the compact object than the GeV zone.

Another important issue is what happens to the relativistic protons once the confinement condition for 10~PeV protons is no longer satisfied downstream of the compact PeV emission zone. The detected emission can be produced either at the proton acceleration site itself or, if the acceleration occurs in an inner region where the jet is not yet transparent to $\gamma$ rays, at the distance where the jet becomes transparent. To preserve a strongly orbital-phase-dependent signal, a substantial fraction of the proton energy should be dissipated before the particles escape from the jet. In this sense, efficient adiabatic cooling would require the transparency distance to lie well inside the confinement distance, i.e. with $\xi\equiv H_{\tau,1}/H_{\rm conf}$ being significantly below unity. Under the single power-law extrapolation, however, Appendix~\ref{app:xi} gives $0.6\lesssim\xi<1$, implying that the transparency and confinement boundaries remain relatively close and that the distance available for adiabatic cooling before proton escape is limited. In that case, a fraction of the proton power is expected to escape into the stellar wind, potentially adding a more weakly modulated component. On the other hand, Appendix~\ref{app:xi} also shows that this conclusion can be relaxed if the magnetic-field evolution departs from a single power law: for a broken power-law profile anchored at Point~B, one obtains $\xi\simeq0.2$. This indicates that rapid magnetic-field dissipation remains necessary between the compact PeV zone and the downstream weakly magnetized GeV zone, although the decay in the immediate vicinity of the PeV emission region need not be very steep.

\section{Conclusion}
\label{sec:conclusion}

In this study, we have used the GeV orbital modulation of Cygnus~X-3 as a diagnostic of the physical conditions in the GeV emission zone during the PeV high state observed by LHAASO. By jointly fitting the phase-averaged GeV spectrum and the orbital light curve with an anisotropic external inverse-Compton model, we localized the GeV emission zone to $H_0 \approx 2.8 \times 10^{11}$~cm from the compact object, comparable to the binary orbital separation. The contemporaneous GeV data constrain the statistically allowed region to relatively small values of the magnetic field--size product, with an upper bound of $BH\le 10^{13.3}\,\rm G\, cm$ at the 3$\sigma$ level. Translating the inferred $BH$ constraint into a Hillas limit implies a conservative upper bound of $E_{p,\max}\sim 0.3~{\rm PeV}$ in the GeV emission zone, far below the $\gtrsim 30$~PeV required to account for $\sim 3$--4~PeV photons. This rules out a simple one-zone scenario in which the GeV and PeV emissions originate from the same region.

We further showed that, if the observed PeV photons are produced within the jet, their emission region should lie further inward than the GeV zone. Under a simple single power-law, flux-freezing-like magnetic-field evolution profile ($B\propto H^{-\delta}$ with $1\leq \delta \le 2$), one cannot simultaneously satisfy PeV-photon transparency and the proton-confinement requirement relevant for the observed multi-PeV emission. Scanning all acceptable $3\sigma$ GeV-zone solutions, we find a global minimum $\delta_{\mathrm{min}}>2$, and the corresponding viable region in the illustrative Point~B case appears only for $\delta=3$, at $H/H_0\lesssim0.15$. This inference is also consistent with the geometric indication from the observed GeV--PeV phase lag. Appendix~\ref{app:xi} further shows that, under a single power-law extrapolation, $0.58\lesssim\xi<1$, where $\xi\equiv H_{\tau,1}/H_{\rm conf}$, implying limited room for adiabatic cooling before proton escape. However, if the magnetic-field evolution departs from a single power law, the constraint on $\xi$ can be substantially relaxed; for a broken power-law profile anchored at Point~B, one obtains $\xi\simeq0.2$. Together, these results indicate that rapid magnetic-field dissipation remains necessary between the compact PeV zone and the downstream GeV zone, although the decay in the immediate vicinity of the PeV emission region need not be very steep and the strongest dissipation may instead occur farther downstream.

We note that several simplifications made in our calculation may affect the quantitative values but are unlikely to change the qualitative conclusion that the GeV zone cannot host a 30~PeV accelerator. First, our GeV modeling assumes a one-zone steady-state leptonic emitter. Considering an extension of the GeV emission zone along the jet axis \citep{2018MNRAS.481.1455K} or a broader jet opening angle \footnote{A larger opening angle of $\alpha=12^\circ\sim24^\circ$ was reported during minor flares of Cygnus~X-3 \citep{2022MNRASSpencer}.} could broaden the allowed GeV-zone parameter region in the $(B,H)$ plane, but the orbital-modulation requirement still strongly limits the SSC contribution and therefore constrains $B$ to remain low on binary scales. Second, we omit the bulk Lorentz factor in the Hillas criterion, but this would not materially change the inferred $E_{p,\rm max}$ because the jet is unlikely to be highly relativistic; hence the allowed $E_{p,\rm max}$ in the GeV zone remains far below 30~PeV. Finally, we consider a simple conical geometry for the jet. Interactions between jet and wind from the companion star may potentially cause bending or precession of the jet at the orbital plane \citep{Yoon2015, Bosch-Ramon2016}, thereby imprinting orbital modulation in the SSC component. However, detailed modeling shows that the required bending angle under this scenario substantially exceeds the theoretically expected value for plausible wind and jet parameters \citep{2024ApJDmytriiev}, so we do not expect such bending effects to significantly alter our conclusion.

Overall, our results support a picture in which the GeV emission is produced on the binary-scale jet under weak magnetization, whereas the observed PeV photons must arise from a more compact inner region. Protons of $\sim 10~$PeV may be accelerated at the PeV emission zone or even in the vicinity of the central compact object. In this picture, the magnetic field near the PeV zone can remain strong enough to confine multi-PeV protons, while stronger dissipation occurs farther downstream before the flow reaches the GeV emission zone. Multiwavelength timing and spectral modeling of Cygnus~X-3 therefore provides a direct way to probe extreme particle acceleration in compact-object jets and may help clarify the origin of Galactic CRs beyond the knee.

\begin{acknowledgements}
This work is funded by National Natural Science Foundation of China under grants No.~12393852 and 12333006, and Basic Research Program of Jiangsu under grant No.~BK20250059. 
\end{acknowledgements}

\appendix

\renewcommand{\thefigure}{A\arabic{figure}}
\setcounter{figure}{0}

\section{Impact of Eccentricity}
\label{app:Eccentricity}

\begin{figure*}[t]
    \centering
    \includegraphics[width=0.5\textwidth]{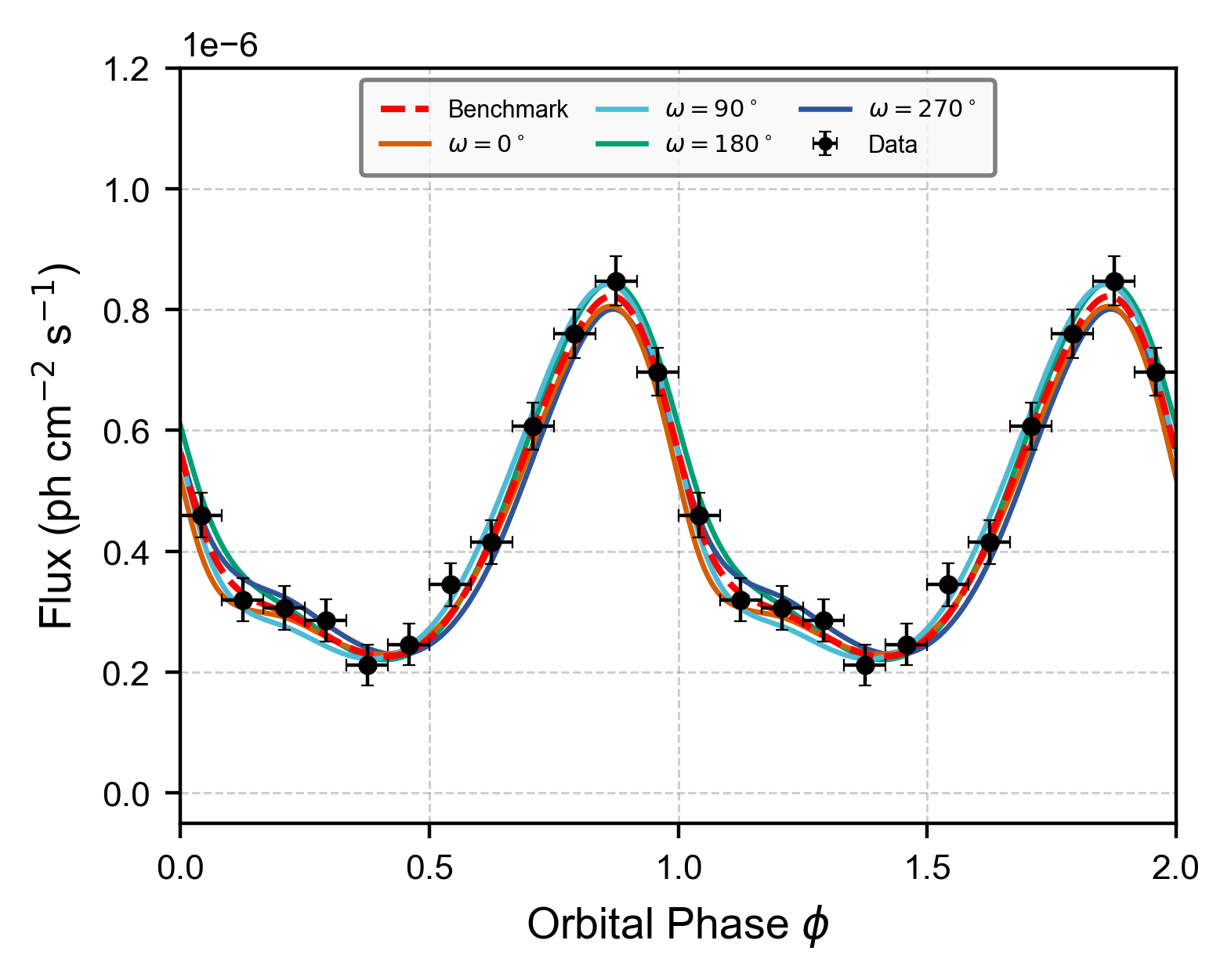}
    \caption{Impact of a small orbital eccentricity ($e \approx 0.03$) on the modeled GeV orbital light curves. The red dashed thick curve represents the circular-orbit best-fit solution. The solid curves in different colors show elliptical-orbit models with four representative arguments of periastron, $\omega = 0^\circ$, $90^\circ$, $180^\circ$, and $270^\circ$. The black circles show the \textit{Fermi}-LAT data. All elliptical cases yield somewhat worse fits than the circular benchmark, indicating that even a small eccentricity introduces a noticeable but not decisive distortion of the orbital modulation profile.}
    \label{fig:elliptical}
\end{figure*}

Long-term monitoring has suggested that Cygnus~X-3 may possess a small but non-zero orbital eccentricity of $e \approx 0.03$. However, this possibility has not been widely adopted in subsequent studies, and the circular-orbit approximation remains the standard choice in most previous works. We therefore treat the eccentric case as a robustness check and assess its impact by replacing the circular orbit with a Keplerian elliptical orbit. We test four representative arguments of periastron, $\omega = 0^\circ$, $90^\circ$, $180^\circ$, and $270^\circ$, and the resulting model light curves are shown in Fig.~\ref{fig:elliptical}.

Relative to the circular-orbit best-fit model, all elliptical cases yield larger chi-square values, namely $\chi^2 = 9.5$, 8.0, 7.3, and 9.9 for $\omega = 0^\circ$, $90^\circ$, $180^\circ$, and $270^\circ$, respectively, compared with $\chi^2_{\mathrm{min}} = 4$ for the circular-orbit best-fit solution. This increase indicates that even a modest eccentricity, $e \approx 0.03$, has a visible effect on the modeled GeV orbital light curve.

In our calculation, the eccentricity affects the light curve in several coupled ways. First, it modifies the mapping between orbital phase and true anomaly, introducing non-uniform orbital motion. Second, the instantaneous binary separation varies with phase, which changes the companion-photon energy density and the anisotropic EC scattering geometry. Third, because the IC cooling rate depends on the local stellar radiation field, the steady-state electron distribution and hence the phase-resolved EC and SSC fluxes are also altered. Therefore, the effect of $e \neq 0$ is not limited to a simple geometric phase shift, but instead produces a systematic reshaping of the orbital modulation profile.

The model with $\omega=0^\circ$ already shows a noticeable degradation in fit quality relative to the circular-orbit benchmark. By contrast, the chi-square values obtained for the four tested arguments of periastron do not show significant differences from one another. This indicates that the main source of the chi-square increase is the eccentric modulation itself, rather than any particular orientation of periastron. In other words, the present \textit{Fermi}-LAT orbital light curve appears to favor a nearly symmetric modulation profile, which is more naturally reproduced in the circular-orbit approximation. Nevertheless, the degradation remains moderate, and the inferred physical picture is unchanged. In particular, the GeV emission is still dominated by anisotropic EC scattering, and all of the main conclusions of this work remain valid when a small eccentricity is taken into account.

\section{The fitting results for Point A}
\label{app:ABcompare}

\begin{figure*}[t]
    \centering
    \includegraphics[width=0.45\textwidth]{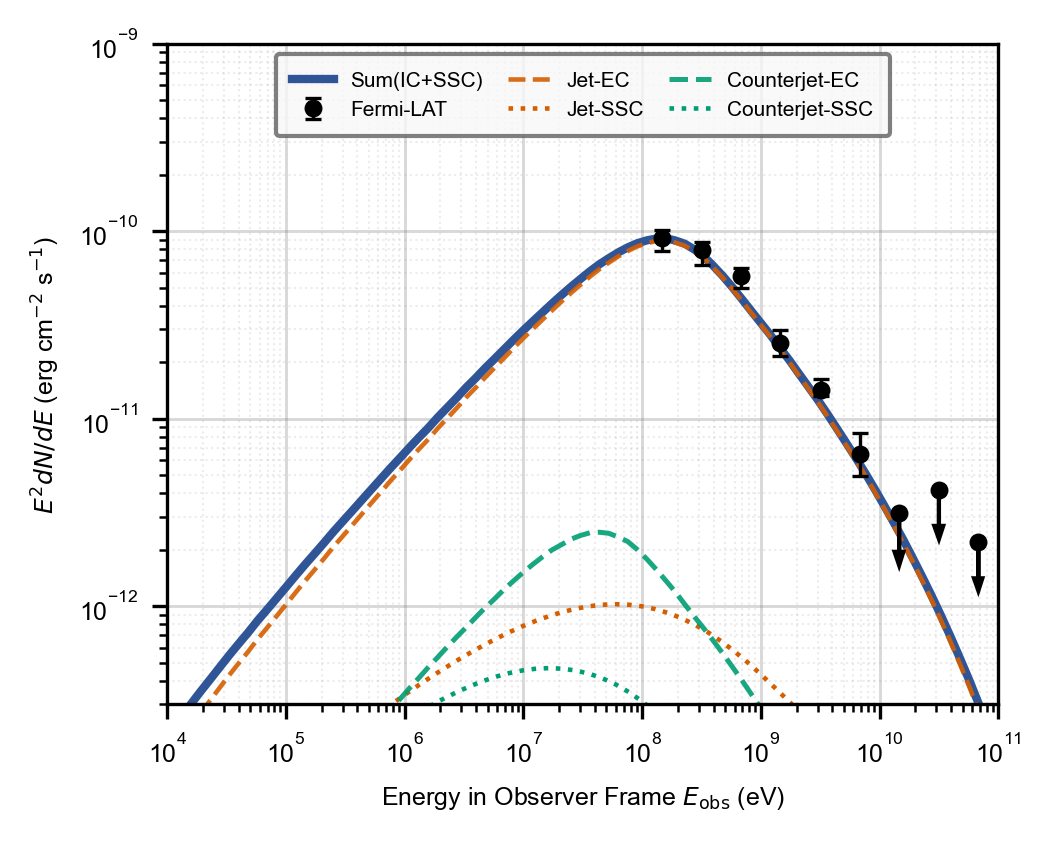}
    \includegraphics[width=0.45\textwidth]{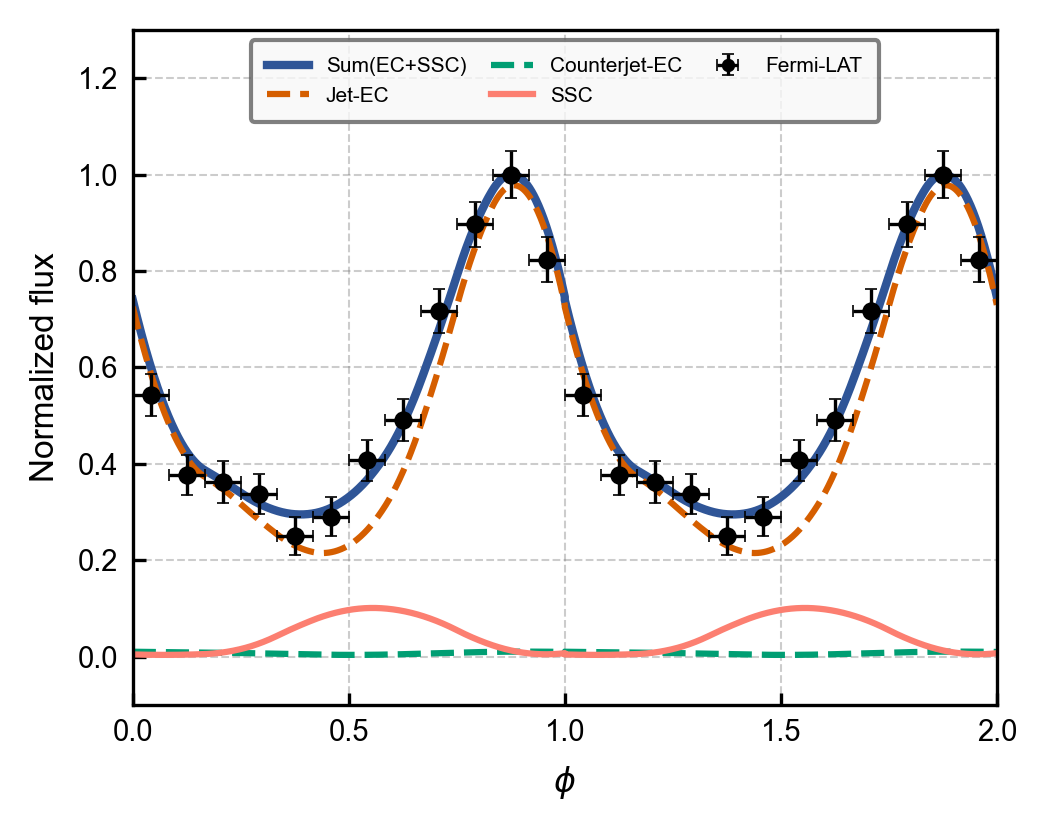}
    \caption{Best-fit GeV spectrum and orbital light curve for Point A. Left: phase-averaged GeV spectrum. Right: corresponding orbital light curve.}
    \label{fig:ABcompare}
\end{figure*}

Point A in Figure~\ref{fig:map} marks the largest-$BH$ solution formally enclosed by the $3\sigma$ contour. Its best-fit spectrum and orbital light curve are shown in Figure~\ref{fig:ABcompare}. The fitted parameters are
$H_0=1.8\times10^{11}\ {\rm cm},\,\beta_j=0.62,\,\phi_j=200^\circ,\,\theta_j=57^\circ,\,p=3.45,\,K_{\rm inj}=1.6\times10^{47}\ {\rm s^{-1}},\,\gamma_1=1770,\,B_0=150\ {\rm G},$
with a minimum chi-square of $\chi^2_{\rm min}=15.8$, with equal contributions from the spectrum and the orbital light curve ($\chi^2_{\rm spec}=7.9$, $\chi^2_{\rm lc}=7.9$).

Point A is characterized by a much stronger magnetic field and a smaller distance from the compact object than the best-fit GeV-zone solution. In this case, the enhanced synchrotron photon density makes the SSC contribution non-negligible, so the GeV spectrum is no longer described by a purely EC-dominated shape. In particular, the additional SSC contribution tends to fill the low-energy LAT band, while the stronger cooling suppresses the highest-energy electrons more efficiently. Point A is therefore useful as an extreme case with the largest allowed $BH$.

\section{Phase Lag}
\label{app:ratio}

The observed orbital phase lag between the PeV and GeV light curves provides an additional geometric clue to the relative locations of the two emission zones. If the two emission regions are separated mainly along the tilted jet, the corresponding phase offset is dominated by geometric projection. Writing the orbital modulation in terms of the phase $\phi$, the geometric contribution can be approximated as
$\Delta \phi_{\mathrm{geom}} \simeq \frac{\Delta H \sin \theta_{\mathrm{j}}}{2\pi a},$
where $\Delta H = H_0 - H_1 = H_0(1-\zeta)$ and $\zeta \equiv H_1/H_0$. This gives
\begin{equation}
\zeta \simeq 1 - \frac{2\pi a\,\Delta \phi_{\mathrm{geom}}}{H_0 \sin \theta_{\mathrm{j}}}.
\label{equ:ratio}
\end{equation}
Using the observed GeV--PeV phase offset of roughly $\Delta\phi \sim 0.08$--$0.28$ and the best-fit GeV-zone parameters, $H_0\simeq2.8\times10^{11}$~cm and $\theta_{\mathrm{j}}\simeq46^\circ$, we obtain a rough geometric upper bound of $\zeta \lesssim 0.3$. This indicates that the PeV emission region is located significantly closer to the compact object than the GeV emission zone.

\begin{figure}[htbp]
    \centering
    \includegraphics[width=0.5\textwidth]{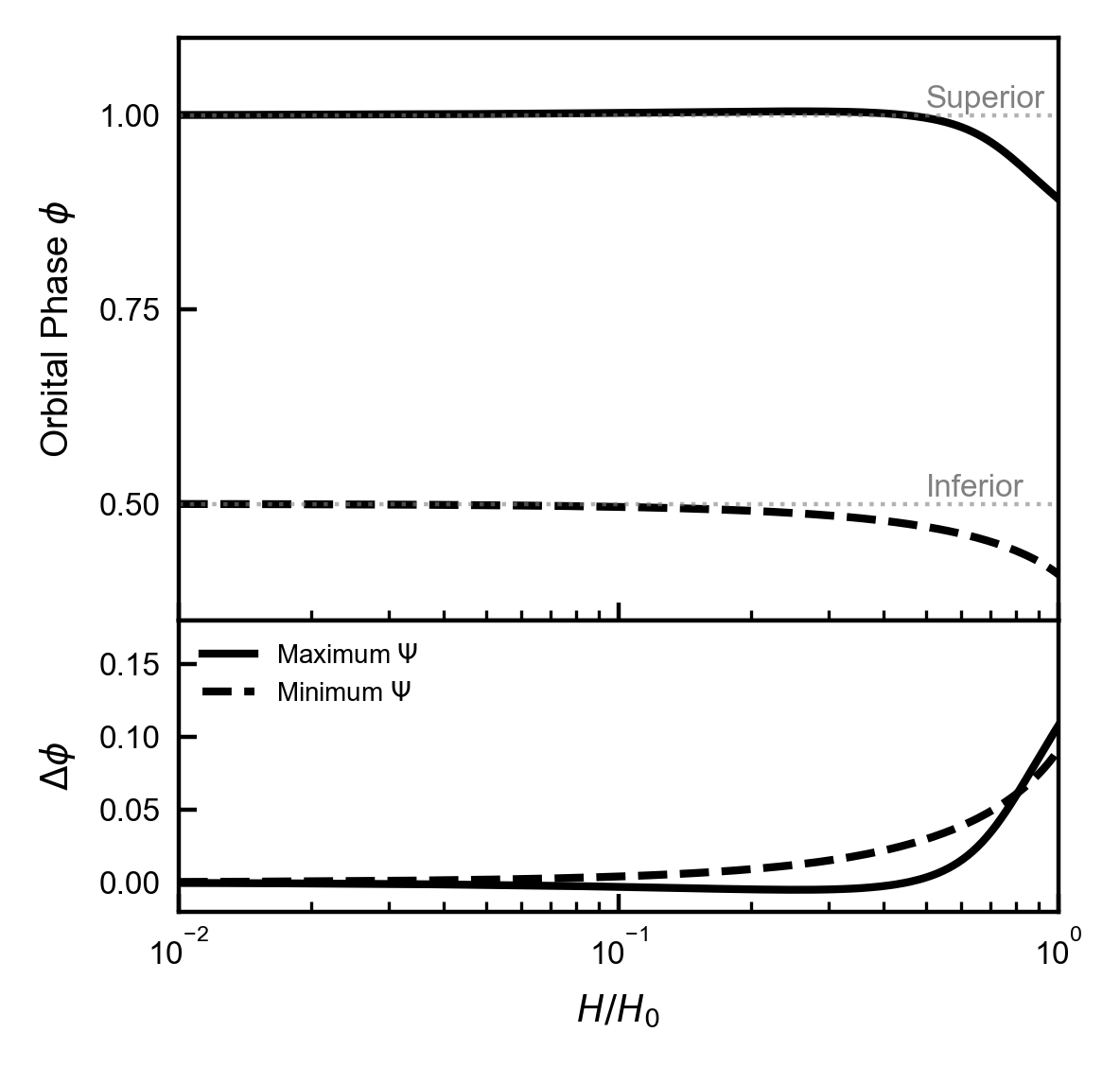}
    \caption{\textbf{Upper panel:} Evolution of the orbital phases corresponding to the maximum (solid line) and minimum (dashed line) EC scattering efficiency as functions of $H/H_0$. The horizontal dotted lines mark the superior and inferior conjunction phases. \textbf{Lower panel:} Phase offsets of the EC extrema relative to the conjunctions. The solid curve shows $\Delta\phi = 1.0 - \phi_{\mathrm{max}}$, and the dashed curve shows $\Delta\phi = 0.5 - \phi_{\mathrm{min}}$. As the emission zone moves inward ($H/H_0 \ll 1$), both offsets approach zero, indicating that a compact inner emitter naturally peaks closer to the conjunction phases.}
    \label{fig:phase}
\end{figure}

Next, we examine the orbital modulation of the EC emission. The interaction efficiency scales with the collision angle $\Psi$ between the target photon unit vector $\boldsymbol{e}_*$ and the line-of-sight vector $\boldsymbol{e}_{\mathrm{obs}}$. The cosine of this scattering angle is geometrically given by
\begin{equation}
\cos\Psi(\theta) = \boldsymbol{e}_* \cdot \boldsymbol{e}_{\mathrm{obs}} = \frac{-a \sin i \cos \theta + H\left(\cos i \cos \theta_{\mathrm{j}} - \sin i \cos \phi_{\mathrm{j}} \sin \theta_{\mathrm{j}}\right)}{R(\theta)}.
\label{eq:cospsi}
\end{equation}
 The orbital phases of maximum and minimum flux, denoted as $\phi_{\mathrm{max}}$ and $\phi_{\mathrm{min}}$, correspond to the extrema of the scattering efficiency. By numerically solving the condition $\partial \cos\Psi(\theta) / \partial \theta = 0$ for varying emission distance $H$ and mapping the solutions back to the orbital phase $\phi$, we track the evolution of the peak phase. Figure~\ref{fig:phase} illustrates $\phi_{\mathrm{max}}$ (solid curve) and $\phi_{\mathrm{min}}$ (dashed curve) as functions of $H/H_0$. The geometric parameters are fixed to the updated best-fit values derived for the GeV emission zone, namely $H_0 \simeq 2.8\times10^{11}$~cm, $\theta_{\mathrm{j}} \simeq 46^\circ$, and $\phi_{\mathrm{j}} \simeq 196^\circ$. In the compact regime ($H/H_0 \ll 1$), the emission extrema asymptotically align with the superior ($\phi \approx 0.0$ or $1.0$) and inferior ($\phi \approx 0.5$) conjunctions. As the height of the emission zone increases toward the GeV zone, geometric parallax shifts the extrema away from the conjunctions. For $H \sim H_0$, we obtain $\phi_{\mathrm{max}} \approx 0.91$ and $\phi_{\mathrm{min}} \approx 0.43$, corresponding to phase offsets of order $\Delta\phi \sim 0.07$--$0.09$ from the conjunction phases. Therefore, if the PeV emission originates from a more compact inner region ($H_1 < H_0$), its orbital peak is naturally expected to move closer to the conjunction phase. This geometric trend is consistent with the inference that the PeV emission zone lies inside the GeV zone.

\section{The ratio of transparency to confinement distances}
\label{app:xi}

We define two characteristic distances in the inner PeV zone: the confinement distance $H_{\rm conf}$, defined by
$E_{\rm conf}(H_{\rm conf})=eB(H_{\rm conf})\alpha H_{\rm conf}=10~\mathrm{PeV}$,
where $10~\mathrm{PeV}$ is adopted here as a robust threshold, and the transparency distance $H_{\tau,1}$, defined by $\tau_{\gamma B}(H_{\tau,1})=1$. Their ratio, $\xi \equiv H_{\tau,1}/H_{\rm conf}$, measures the relative locations of the transparency and confinement boundaries.

Using Equations~(\ref{equ:BH}) and (\ref{equ:tau}), $\xi$ is fully determined by $(B_0,H_0,\delta)$:
\begin{equation}
\xi(B_0,H_0,\delta)
=
\frac{
\left[
\frac{\delta-1}{\delta D}
W\!\left(
\frac{\delta D}{\delta-1}
A^{\delta/(\delta-1)}
\right)
\right]^{1/\delta}
}{
\left({eB_0H_0\alpha}/{10~{\rm PeV}}\right)^{1/(\delta-1)}
},
\label{eq:xi2_app}
\end{equation}
where $W$ is the Lambert $W$ function. Under the single power-law extrapolation, scanning the $(B_0,H_0)$ parameter space within the $3\sigma$ contour for $\delta>2$ gives $0.6\lesssim \xi < 1$. Thus, in the single-PL picture, the transparency and confinement boundaries remain relatively close to each other.

This result, however, relies on assuming that the magnetic field evolves as a single power law from the GeV zone down to the inner PeV zone. As an illustrative alternative, we also consider a broken power-law magnetic-field profile,
\begin{equation}
B(H)=
\begin{cases}
B_0 \left(\dfrac{H_0}{H}\right)^{\delta_{\rm out}}, & H_b \le H \le H_0, \\
B_0 \left(\dfrac{H_0}{H_b}\right)^{\delta_{\rm out}}
\left(\dfrac{H_b}{H}\right)^{\delta_{\rm in}}, & H < H_b,
\end{cases}
\label{eq:brokenB}
\end{equation}
where $(B_0,H_0)$ denotes the GeV-zone reference point, $H_b$ is the break distance, and $\delta_{\rm out}$ and $\delta_{\rm in}$ are the outer and inner decay indices, respectively.

Taking the Point~B reference point, $(B_0,H_0)=(8.2~{\rm G},\,1.1\times10^{12}~{\rm cm})$, and adopting $\delta_{\rm out}=3$, $\delta_{\rm in}=1$, and $H_b/H_0=0.15$, we obtain $H_{\rm conf}\simeq1.54\times10^{-1}H_0$ and $H_{\tau,1}\simeq3.03\times10^{-2}H_0$, so that $\xi\simeq0.2$. Thus, once the magnetic-field evolution departs from a single power law, the constraint on $\xi$ can be substantially relaxed. In particular, a steeper outer decay generally leads to a smaller $\xi$; for example, $\delta_{\rm out}\sim4$ can yield $\xi\sim0.1$. Therefore, even if the magnetic-field profile departs from a single power law, rapid magnetic-field dissipation remains necessary between the compact PeV zone and the downstream GeV zone. What the broken-power-law profile changes is not this requirement itself, but only where the strongest dissipation occurs: the field in the immediate vicinity of the PeV emission zone need not decay very steeply, so that high-energy protons can remain confined within the jet, whereas the stronger dissipation can take place farther downstream toward the weakly magnetized GeV zone.

\bibliography{sample631}{}

@article{LHAASO2025_microquasar,
    author = {{LHAASO Collaboration}},
    title = {Ultrahigh-energy gamma-ray emission associated with black hole–jet systems},
    journal = {National Science Review},
    volume = {12},
    number = {12},
    pages = {nwaf496},
    year = {2025},
    month = {11},
    issn = {2095-5138},
    doi = {10.1093/nsr/nwaf496},
    url = {https://doi.org/10.1093/nsr/nwaf496},
    archivePrefix = {arXiv},
       eprint = {2410.08988},
 primaryClass = {astro-ph.HE},
}

@ARTICLE{Hillas1984,
       author = {{Hillas}, A.~M.},
        title = "{The Origin of Ultra-High-Energy Cosmic Rays}",
      journal = {\araa},
         year = 1984,
        month = jan,
       volume = {22},
        pages = {425-444},
          doi = {10.1146/annurev.aa.22.090184.002233},
       adsurl = {https://ui.adsabs.harvard.edu/abs/1984ARA&A..22..425H}
}

@ARTICLE{Yoon2015,
       author = {{Yoon}, D. and {Heinz}, S.},
        title = "{Global Simulations of the Interaction of Microquasar Jets with a Stellar Wind in High-mass X-ray Binaries}",
      journal = {\apj},
         year = 2015,
        month = mar,
       volume = {801},
       number = {1},
          eid = {55},
        pages = {55},
          doi = {10.1088/0004-637X/801/1/55},
archivePrefix = {arXiv},
       eprint = {1501.03827},
 primaryClass = {astro-ph.HE},
       adsurl = {https://ui.adsabs.harvard.edu/abs/2015ApJ...801...55Y}
}

@ARTICLE{Wei2025,
       author = {{Wei}, Yu-Jia and {Murase}, Kohta and {Zhang}, B. Theodore},
        title = "{Unveiling Multimessenger Emission from Hidden Cores of Microquasars}",
      journal = {arXiv e-prints},
         year = 2025,
        month = dec,
          eid = {arXiv:2512.23231},
        pages = {arXiv:2512.23231},
          doi = {10.48550/arXiv.2512.23231},
archivePrefix = {arXiv},
       eprint = {2512.23231},
 primaryClass = {astro-ph.HE},
       adsurl = {https://ui.adsabs.harvard.edu/abs/2025arXiv251223231W}
}

@ARTICLE{Bosch-Ramon2016,
       author = {{Bosch-Ramon}, V. and {Barkov}, M.~V.},
        title = "{The effects of the stellar wind and orbital motion on the jets of high-mass microquasars}",
      journal = {\aap},
         year = 2016,
        month = may,
       volume = {590},
          eid = {A119},
        pages = {A119},
          doi = {10.1051/0004-6361/201628564},
archivePrefix = {arXiv},
       eprint = {1604.06360},
 primaryClass = {astro-ph.HE},
       adsurl = {https://ui.adsabs.harvard.edu/abs/2016A&A...590A.119B}
}

@ARTICLE{WangJS2018,
       author = {{Wang}, Jie-Shuang and {Liu}, Ruo-Yu and {Aharonian}, Felix and {Dai}, Zi-Gao},
        title = "{Analytical treatment for the development of electromagnetic cascades in intense magnetic fields}",
      journal = {\prd},
         year = 2018,
        month = may,
       volume = {97},
       number = {10},
          eid = {103016},
        pages = {103016},
          doi = {10.1103/PhysRevD.97.103016},
archivePrefix = {arXiv},
       eprint = {1805.03040},
 primaryClass = {astro-ph.HE},
       adsurl = {https://ui.adsabs.harvard.edu/abs/2018PhRvD..97j3016W}
}

@ARTICLE{Alfaro2024,
       author = {{Alfaro}, R. and {Alvarez}, C. and {Arteaga-Vel{\'a}zquez}, J.~C. and {Avila Rojas}, D. and {Ayala Solares}, H.~A. and {Babu}, R. and {Belmont-Moreno}, E. and {Caballero-Mora}, K.~S. and {Capistr{\'a}n}, T. and {Carrami{\~n}ana}, A. and {Casanova}, S. and {Cotti}, U. and {Cotzomi}, J. and {Couti{\~n}o de Le{\'o}n}, S. and {De la Fuente}, E. and {Depaoli}, D. and {Di Lalla}, N. and {Diaz Hernandez}, R. and {Dingus}, B.~L. and {DuVernois}, M.~A. and {Durocher}, M. and {D{\'\i}az-V{\'e}lez}, J.~C. and {Engel}, K. and {Espinoza}, C. and {Fan}, K.~L. and {Fang}, K. and {Fraija}, N. and {Fraija}, S. and {Garc{\'\i}a-Gonz{\'a}lez}, J.~A. and {Garfias}, F. and {Gonzalez Mu{\~n}oz}, A. and {Gonz{\'a}lez}, M.~M. and {Goodman}, J.~A. and {Groetsch}, S. and {Harding}, J.~P. and {Herzog}, I. and {Hinton}, J. and {Huang}, D. and {Hueyotl-Zahuantitla}, F. and {H{\"u}ntemeyer}, P. and {Iriarte}, A. and {Joshi}, V. and {Kaufmann}, S. and {Kieda}, D. and {de Le{\'o}n}, C. and {Lee}, J. and {Le{\'o}n Vargas}, H. and {Linnemann}, J.~T. and {Longinotti}, A.~L. and {Luis-Raya}, G. and {Malone}, K. and {Martinez}, O. and {Mart{\'\i}nez-Castro}, J. and {Matthews}, J.~A. and {Miranda-Romagnoli}, P. and {Morales-Soto}, J.~A. and {Moreno}, E. and {Mostaf{\'a}}, M. and {Nayerhoda}, A. and {Nellen}, L. and {Newbold}, M. and {Nisa}, M.~U. and {Noriega-Papaqui}, R. and {Olivera-Nieto}, L. and {Omodei}, N. and {Osorio}, M. and {P{\'e}rez Araujo}, Y. and {P{\'e}rez-P{\'e}rez}, E.~G. and {Rho}, C.~D. and {Rosa-Gonz{\'a}lez}, D. and {Ruiz-Velasco}, E. and {Salazar}, H. and {Salazar-Gallegos}, D. and {Sandoval}, A. and {Schneider}, M. and {Serna-Franco}, J. and {Smith}, A.~J. and {Son}, Y. and {Springer}, R.~W. and {Tibolla}, O. and {Tollefson}, K. and {Torres}, I. and {Torres-Escobedo}, R. and {Turner}, R. and {Ure{\~n}a-Mena}, F. and {Varela}, E. and {Villase{\~n}or}, L. and {Wang}, X. and {Watson}, I.~J. and {Willox}, E. and {Yun-C{\'a}rcamo}, S. and {Zhou}, H.},
        title = "{Ultra-high-energy gamma-ray bubble around microquasar V4641 Sgr}",
      journal = {\nat},
         year = 2024,
        month = oct,
       volume = {634},
       number = {8034},
        pages = {557-560},
          doi = {10.1038/s41586-024-07995-9},
archivePrefix = {arXiv},
       eprint = {2410.16117},
 primaryClass = {astro-ph.HE},
       adsurl = {https://ui.adsabs.harvard.edu/abs/2024Natur.634..557A}
}

@BOOK{1964ocrbookG,
  author    = {Ginzburg, V. L. and Syrovatskii, S. I.},
  title     = {The Origin of Cosmic Rays},
  year      = {1964},
  publisher = {Pergamon Press},
  address   = {Oxford},
  adsurl    = {https://ui.adsabs.harvard.edu/abs/1964ocr..book.....G}
}

@ARTICLE{2024SciBu..69..449L,
       author = {{LHAASO Collaboration}},
        title = "{An ultrahigh-energy {\ensuremath{\gamma}} -ray bubble powered by a super PeVatron}",
      journal = {Science Bulletin},
         year = 2024,
        month = feb,
       volume = {69},
       number = {4},
        pages = {449-457},
          doi = {10.1016/j.scib.2023.12.040},
archivePrefix = {arXiv},
       eprint = {2310.10100},
 primaryClass = {astro-ph.HE},
       adsurl = {https://ui.adsabs.harvard.edu/abs/2024SciBu..69..449L}
}

@ARTICLE{2025arXivLHAASO,
       author = {{The LHAASO Collaboration} and {Cao}, Zhen and {Aharonian}, F. and {Bai}, Y.~X. and {Bao}, Y.~W. and {Bastieri}, D. and {Bi}, X.~J. and {Bi}, Y.~J. and {Bian}, W. and {Blunier}, J. and {Bukevich}, A.~V. and {Cai}, C.~M. and {Cai}, Y.~Y. and {Cao}, W.~Y. and {Cao}, Zhe and {Chang}, J. and {Chang}, J.~F. and {Chen}, E.~S. and {Chen}, G.~H. and {Chen}, H.~K. and {Chen}, L.~F. and {Chen}, Liang and {Chen}, Long and {Chen}, M.~J. and {Chen}, M.~L. and {Chen}, Q.~H. and {Chen}, S. and {Chen}, S.~H. and {Chen}, S.~Z. and {Chen}, T.~L. and {Chen}, X.~B. and {Chen}, X.~J. and {Chen}, X.~P. and {Chen}, Y. and {Cheng}, N. and {Cheng}, Q.~Y. and {Cheng}, Y.~D. and {Cui}, M.~Y. and {Cui}, S.~W. and {Cui}, X.~H. and {Cui}, Y.~D. and {Dai}, B.~Z. and {Dai}, H.~L. and {Dai}, Z.~G. and {Danzengluobu} and {Diao}, Y.~X. and {Dong}, A.~J. and {Dong}, X.~Q. and {Duan}, K.~K. and {Fan}, J.~H. and {Fan}, Y.~Z. and {Fang}, J. and {Fang}, J.~H. and {Fang}, K. and {Feng}, C.~F. and {Feng}, H. and {Feng}, L. and {Feng}, S.~H. and {Feng}, X.~T. and {Feng}, Y. and {Feng}, Y.~L. and {Gabici}, S. and {Gao}, B. and {Gao}, Q. and {Gao}, W. and {Gao}, W.~K. and {Ge}, M.~M. and {Ge}, T.~T. and {Geng}, L.~S. and {Giacinti}, G. and {Gong}, G.~H. and {Gou}, Q.~B. and {Gu}, M.~H. and {Guo}, F.~L. and {Guo}, J. and {Guo}, K.~J. and {Guo}, X.~L. and {Guo}, Y.~Q. and {Guo}, Y.~Y. and {Han}, R.~P. and {Hannuksela}, O.~A. and {Hasan}, M. and {He}, H.~H. and {He}, H.~N. and {He}, J.~Y. and {He}, X.~Y. and {He}, Y. and {Hern{\'a}ndez-Cadena}, S. and {Hou}, B.~W. and {Hou}, C. and {Hou}, X. and {Hu}, H.~B. and {Hu}, S.~C. and {Huang}, C. and {Huang}, D.~H. and {Huang}, J.~J. and {Huang}, X.~L. and {Huang}, X.~T. and {Huang}, X.~Y. and {Huang}, Y. and {Huang}, Y.~Y. and {Inventar}, A. and {Ji}, X.~L. and {Jia}, H.~Y. and {Jia}, K. and {Jiang}, H.~B. and {Jiang}, K. and {Jiang}, X.~W. and {Jiang}, Z.~J. and {Jin}, M. and {Kaci}, S. and {Kang}, M.~M. and {Karpikov}, I. and {Khangulyan}, D. and {Kuleshov}, D. and {Kurinov}, K. and {Li}, Cheng and {Li}, Cong and {Li}, D. and {Li}, F. and {Li}, H.~B. and {Li}, H.~C. and {Li}, Jian and {Li}, Jie and {Li}, K. and {Li}, L. and {Li}, R.~L. and {Li}, S.~D. and {Li}, T.~Y. and {Li}, W.~L. and {Li}, X.~R. and {Li}, Xin and {Li}, Y. and {Li}, Zhe and {Li}, Zhuo and {Liang}, E.~W. and {Liang}, Y.~F. and {Lin}, S.~J. and {Liu}, B. and {Liu}, C. and {Liu}, D. and {Liu}, D.~B. and {Liu}, H. and {Liu}, J. and {Liu}, J.~L. and {Liu}, J.~R. and {Liu}, M.~Y. and {Liu}, R.~Y. and {Liu}, S.~M. and {Liu}, W. and {Liu}, X. and {Liu}, Y. and {Liu}, Y. and {Liu}, Y.~N. and {Lou}, Y.~Q. and {Luo}, Q. and {Luo}, Y. and {Lv}, H.~K. and {Ma}, B.~Q. and {Ma}, L.~L. and {Ma}, X.~H. and {Maliy}, I.~O. and {Mao}, J.~R. and {Min}, Z. and {Mitthumsiri}, W. and {Mizuno}, Y. and {Mou}, G.~B. and {Neronov}, A. and {Ng}, K.~C.~Y. and {Ni}, M.~Y. and {Nie}, L. and {Ou}, L.~J. and {Ou}, Z.~W. and {Pattarakijwanich}, P. and {Pei}, Z.~Y. and {Peng}, D.~Y. and {Qi}, J.~C. and {Qi}, M.~Y. and {Qin}, J.~J. and {Qu}, D. and {Raza}, A. and {Ren}, C.~Y. and {Ruffolo}, D. and {S{\'a}iz}, A. and {Savchenko}, D. and {Semikoz}, D. and {Shao}, L. and {Shchegolev}, O. and {Shen}, Y.~Z. and {Sheng}, X.~D. and {Shi}, Z.~D. and {Shu}, F.~W. and {Song}, H.~C. and {Stenkin}, Yu. V. and {Stepanov}, V. and {Su}, Y. and {Sun}, D.~X. and {Sun}, H. and {Sun}, J.~X. and {Sun}, Q.~N.},
        title = "{Cygnus X-3: A variable petaelectronvolt gamma-ray source}",
      journal = {arXiv e-prints},
         year = 2025,
        month = dec,
          eid = {arXiv:2512.16638},
        pages = {arXiv:2512.16638},
          doi = {10.48550/arXiv.2512.16638},
archivePrefix = {arXiv},
       eprint = {2512.16638},
 primaryClass = {astro-ph.HE},
       adsurl = {https://ui.adsabs.harvard.edu/abs/2025arXiv251216638T}
}

@ARTICLE{1967ApJ...148L.119G,
       author = {{Giacconi}, R. and {Gorenstein}, P. and {Gursky}, H. and {Waters}, J.~R.},
        title = "{An X-Ray Survey of the Cygnus Region}",
      journal = {\apjl},
         year = 1967,
        month = jun,
       volume = {148},
        pages = {L119},
          doi = {10.1086/180028},
       adsurl = {https://ui.adsabs.harvard.edu/abs/1967ApJ...148L.119G}
}

@ARTICLE{2008MNRAS.384..278H,
       author = {{Hjalmarsdotter}, L. and {Zdziarski}, A.~A. and {Larsson}, S. and {Beckmann}, V. and {McCollough}, M. and {Hannikainen}, D.~C. and {Vilhu}, O.},
        title = "{The nature of the hard state of Cygnus X-3}",
      journal = {\mnras},
         year = 2008,
        month = feb,
       volume = {384},
       number = {1},
        pages = {278-290},
          doi = {10.1111/j.1365-2966.2007.12688.x},
archivePrefix = {arXiv},
       eprint = {0707.2032},
 primaryClass = {astro-ph},
       adsurl = {https://ui.adsabs.harvard.edu/abs/2008MNRAS.384..278H}
}

@ARTICLE{2022ApJ...926..123A,
       author = {{Antokhin}, Igor I. and {Cherepashchuk}, Anatol M. and {Antokhina}, Eleonora A. and {Tatarnikov}, Andrey M.},
        title = "{Near-IR and X-Ray Variability of Cyg X-3: Evidence for a Compact IR Source and Complex Wind Structures}",
      journal = {\apj},
         year = 2022,
        month = feb,
       volume = {926},
       number = {2},
          eid = {123},
        pages = {123},
          doi = {10.3847/1538-4357/ac4047},
archivePrefix = {arXiv},
       eprint = {2112.04805},
 primaryClass = {astro-ph.HE},
       adsurl = {https://ui.adsabs.harvard.edu/abs/2022ApJ...926..123A}
}

@ARTICLE{2004ApJMiller,
       author = {{Miller-Jones}, James C.~A. and {Blundell}, Katherine M. and {Rupen}, Michael P. and {Mioduszewski}, Amy J. and {Duffy}, Peter and {Beasley}, Anthony J.},
        title = "{Time-sequenced Multi-Radio Frequency Observations of Cygnus X-3 in Flare}",
      journal = {\apj},
         year = 2004,
        month = jan,
       volume = {600},
       number = {1},
        pages = {368-389},
          doi = {10.1086/379706},
archivePrefix = {arXiv},
       eprint = {astro-ph/0311277},
 primaryClass = {astro-ph},
       adsurl = {https://ui.adsabs.harvard.edu/abs/2004ApJ...600..368M}
}

@INPROCEEDINGS{2009Miller,
       author = {{Miller-Jones}, J.~C.~A. and {Sakari}, C.~M. and {Dhawan}, V. and {Tudose}, V. and {Fender}, R.~P. and {Paragi}, Z. and {Garrett}, M.},
        title = "{The proper motion and changing jet morphology of Cygnus X-3}",
    booktitle = {8th International e-VLBI Workshop},
         year = 2009,
        month = jan,
          eid = {17},
        pages = {17},
          doi = {10.22323/1.082.0017},
archivePrefix = {arXiv},
       eprint = {0909.2589},
 primaryClass = {astro-ph.GA},
       adsurl = {https://ui.adsabs.harvard.edu/abs/2009evlb.confE..17M}
}

@ARTICLE{2022MNRASSpencer,
       author = {{Spencer}, Ralph E. and {Garrett}, Michael and {Bray}, Justin D. and {Green}, David A.},
        title = "{Major and minor flares on Cygnus X-3 revisited}",
      journal = {\mnras},
         year = 2022,
        month = may,
       volume = {512},
       number = {2},
        pages = {2618-2624},
          doi = {10.1093/mnras/stac666},
archivePrefix = {arXiv},
       eprint = {2203.05637},
 primaryClass = {astro-ph.SR},
       adsurl = {https://ui.adsabs.harvard.edu/abs/2022MNRAS.512.2618S}
}

@ARTICLE{2023ApJMiller,
       author = {{Reid}, M.~J. and {Miller-Jones}, J.~C.~A.},
        title = "{On the Distances to the X-Ray Binaries Cygnus X-3 and GRS 1915+105}",
      journal = {\apj},
         year = 2023,
        month = dec,
       volume = {959},
       number = {2},
          eid = {85},
        pages = {85},
          doi = {10.3847/1538-4357/acfe0c},
archivePrefix = {arXiv},
       eprint = {2309.15027},
 primaryClass = {astro-ph.HE},
       adsurl = {https://ui.adsabs.harvard.edu/abs/2023ApJ...959...85R}
}

@ARTICLE{2009ApJ...695.1111L,
       author = {{Ling}, Zhixing and {Zhang}, Shuang Nan and {Tang}, Shichao},
        title = "{Determining the Distance of Cyg X-3 with its X-Ray Dust Scattering Halo}",
      journal = {\apj},
         year = 2009,
        month = apr,
       volume = {695},
       number = {2},
        pages = {1111-1120},
          doi = {10.1088/0004-637X/695/2/1111},
archivePrefix = {arXiv},
       eprint = {0901.2990},
 primaryClass = {astro-ph.HE},
       adsurl = {https://ui.adsabs.harvard.edu/abs/2009ApJ...695.1111L}
}

@ARTICLE{2024NatAsVeledina,
       author = {{Veledina}, Alexandra and {Muleri}, Fabio and {Poutanen}, Juri and {Podgorn{\'y}}, Jakub and {Dov{\v{c}}iak}, Michal and {Capitanio}, Fiamma and {Churazov}, Eugene and {De Rosa}, Alessandra and {Di Marco}, Alessandro and {Forsblom}, Sofia V. and {Kaaret}, Philip and {Krawczynski}, Henric and {La Monaca}, Fabio and {Loktev}, Vladislav and {Lutovinov}, Alexander A. and {Molkov}, Sergey V. and {Mushtukov}, Alexander A. and {Ratheesh}, Ajay and {Rodriguez Cavero}, Nicole and {Steiner}, James F. and {Sunyaev}, Rashid A. and {Tsygankov}, Sergey S. and {Weisskopf}, Martin C. and {Zdziarski}, Andrzej A. and {Bianchi}, Stefano and {Bright}, Joe S. and {Bursov}, Nikolaj and {Costa}, Enrico and {Egron}, Elise and {Garcia}, Javier A. and {Green}, David A. and {Gurwell}, Mark and {Ingram}, Adam and {Kajava}, Jari J.~E. and {Kale}, Ruta and {Kraus}, Alex and {Malyshev}, Denys and {Marin}, Fr{\'e}d{\'e}ric and {Matt}, Giorgio and {McCollough}, Michael and {Mereminskiy}, Ilya A. and {Nizhelsky}, Nikolaj and {Piano}, Giovanni and {Pilia}, Maura and {Pittori}, Carlotta and {Rao}, Ramprasad and {Righini}, Simona and {Soffitta}, Paolo and {Shevchenko}, Anton and {Svoboda}, Jiri and {Tombesi}, Francesco and {Trushkin}, Sergei A. and {Tsybulev}, Peter and {Ursini}, Francesco and {Wu}, Kinwah and {Agudo}, Iv{\'a}n and {Antonelli}, Lucio A. and {Bachetti}, Matteo and {Baldini}, Luca and {Baumgartner}, Wayne H. and {Bellazzini}, Ronaldo and {Bongiorno}, Stephen D. and {Bonino}, Raffaella and {Brez}, Alessandro and {Bucciantini}, Niccol{\`o} and {Castellano}, Simone and {Cavazzuti}, Elisabetta and {Chen}, Chien-Ting and {Ciprini}, Stefano and {Del Monte}, Ettore and {Di Gesu}, Laura and {Di Lalla}, Niccol{\`o} and {Donnarumma}, Immacolata and {Doroshenko}, Victor and {Ehlert}, Steven R. and {Enoto}, Teruaki and {Evangelista}, Yuri and {Fabiani}, Sergio and {Ferrazzoli}, Riccardo and {Gunji}, Shuichi and {Hayashida}, Kiyoshi and {Heyl}, Jeremy and {Iwakiri}, Wataru and {Jorstad}, Svetlana G. and {Karas}, Vladimir and {Kislat}, Fabian and {Kitaguchi}, Takao and {Kolodziejczak}, Jeffery J. and {Latronico}, Luca and {Liodakis}, Ioannis and {Maldera}, Simone and {Manfreda}, Alberto and {Marinucci}, Andrea and {Marscher}, Alan P. and {Marshall}, Herman L. and {Massaro}, Francesco and {Mitsuishi}, Ikuyuki and {Mizuno}, Tsunefumi and {Negro}, Michela and {Ng}, Chi-Yung and {O'Dell}, Stephen L. and {Omodei}, Nicola and {Oppedisano}, Chiara and {Papitto}, Alessandro and {Pavlov}, George G. and {Peirson}, Abel L. and {Perri}, Matteo and {Pesce-Rollins}, Melissa and {Petrucci}, Pierre-Olivier and {Possenti}, Andrea and {Puccetti}, Simonetta and {Ramsey}, Brian D. and {Rankin}, John and {Roberts}, Oliver and {Romani}, Roger W. and {Sgr{\`o}}, Carmelo and {Slane}, Patrick and {Spandre}, Gloria and {Swartz}, Doug and {Tamagawa}, Toru and {Tavecchio}, Fabrizio and {Taverna}, Roberto and {Tawara}, Yuzuru and {Tennant}, Allyn F. and {Thomas}, Nicholas E. and {Trois}, Alessio and {Turolla}, Roberto and {Vink}, Jacco and {Xie}, Fei and {Zane}, Silvia},
        title = "{Cygnus X-3 revealed as a Galactic ultraluminous X-ray source by IXPE}",
      journal = {Nature Astronomy},
         year = 2024,
        month = aug,
       volume = {8},
        pages = {1031-1046},
          doi = {10.1038/s41550-024-02294-9},
archivePrefix = {arXiv},
       eprint = {2303.01174},
 primaryClass = {astro-ph.HE},
       adsurl = {https://ui.adsabs.harvard.edu/abs/2024NatAs...8.1031V}
}

@ARTICLE{2001ApJ...553..766M,
       author = {{Mioduszewski}, Amy J. and {Rupen}, Michael P. and {Hjellming}, Robert M. and {Pooley}, Guy G. and {Waltman}, Elizabeth B.},
        title = "{A One-sided Highly Relativistic Jet from Cygnus X-3}",
      journal = {\apj},
         year = 2001,
        month = jun,
       volume = {553},
       number = {2},
        pages = {766-775},
          doi = {10.1086/320965},
archivePrefix = {arXiv},
       eprint = {astro-ph/0102018},
 primaryClass = {astro-ph},
       adsurl = {https://ui.adsabs.harvard.edu/abs/2001ApJ...553..766M}
}

@ARTICLE{2009SciFermi,
       author = {{Fermi LAT Collaboration} and {Abdo}, A.~A. and {Ackermann}, M. and {Ajello}, M. and {Axelsson}, M. and {Baldini}, L. and {Ballet}, J. and {Barbiellini}, G. and {Bastieri}, D. and {Baughman}, B.~M. and {Bechtol}, K. and {Bellazzini}, R. and {Berenji}, B. and {Blandford}, R.~D. and {Bloom}, E.~D. and {Bonamente}, E. and {Borgland}, A.~W. and {Brez}, A. and {Brigida}, M. and {Bruel}, P. and {Burnett}, T.~H. and {Buson}, S. and {Caliandro}, G.~A. and {Cameron}, R.~A. and {Caraveo}, P.~A. and {Casandjian}, J.~M. and {Cecchi}, C. and {{\c{C}}elik}, {\"O}. and {Chaty}, S. and {Cheung}, C.~C. and {Chiang}, J. and {Ciprini}, S. and {Claus}, R. and {Cohen-Tanugi}, J. and {Cominsky}, L.~R. and {Conrad}, J. and {Corbel}, S. and {Corbet}, R. and {Dermer}, C.~D. and {de Palma}, F. and {Digel}, S.~W. and {do Couto e Silva}, E. and {Drell}, P.~S. and {Dubois}, R. and {Dubus}, G. and {Dumora}, D. and {Farnier}, C. and {Favuzzi}, C. and {Fegan}, S.~J. and {Focke}, W.~B. and {Fortin}, P. and {Frailis}, M. and {Fusco}, P. and {Gargano}, F. and {Gehrels}, N. and {Germani}, S. and {Giavitto}, G. and {Giebels}, B. and {Giglietto}, N. and {Giordano}, F. and {Glanzman}, T. and {Godfrey}, G. and {Grenier}, I.~A. and {Grondin}, M.-H. and {Grove}, J.~E. and {Guillemot}, L. and {Guiriec}, S. and {Hanabata}, Y. and {Harding}, A.~K. and {Hayashida}, M. and {Hays}, E. and {Hill}, A.~B. and {Hjalmarsdotter}, L. and {Horan}, D. and {Hughes}, R.~E. and {Jackson}, M.~S. and {J{\'o}hannesson}, G. and {Johnson}, A.~S. and {Johnson}, T.~J. and {Johnson}, W.~N. and {Kamae}, T. and {Katagiri}, H. and {Kawai}, N. and {Kerr}, M. and {Kn{\"o}dlseder}, J. and {Kocian}, M.~L. and {Koerding}, E. and {Kuss}, M. and {Lande}, J. and {Latronico}, J. and {Lemoine-Goumard}, M. and {Longo}, F. and {Loparco}, F. and {Lott}, B. and {Lovellette}, M.~N. and {Lubrano}, P. and {Madejski}, G.~M. and {Makeev}, A. and {Marchand}, L. and {Marelli}, M. and {Max-Moerbeck}, W. and {Mazziotta}, M.~N. and {McColl}, N. and {McEnery}, J.~E. and {Meurer}, C. and {Michelson}, P.~F. and {Migliari}, S. and {Mitthumsiri}, W. and {Mizuno}, T. and {Monte}, C. and {Monzani}, M.~E. and {Morselli}, A. and {Moskalenko}, I.~V. and {Murgia}, S. and {Nolan}, P.~L. and {Norris}, J.~P. and {Nuss}, E. and {Ohsugi}, T. and {Omodei}, N. and {Ong}, R.~A. and {Ormes}, J.~F. and {Paneque}, D. and {Parent}, D. and {Pelassa}, V. and {Pepe}, M. and {Pesce-Rollins}, M. and {Piron}, F. and {Pooley}, G. and {Porter}, T.~A. and {Pottschmidt}, K. and {Rain{\`o}}, S. and {Rando}, R. and {Ray}, P.~S. and {Razzano}, M. and {Rea}, N. and {Readhead}, A. and {Reimer}, A. and {Reimer}, O. and {Richards}, J.~L. and {Rochester}, L.~S. and {Rodriguez}, J. and {Rodriguez}, A.~Y. and {Romani}, R.~W. and {Ryde}, F. and {Sadrozinski}, H.~F.-W. and {Sander}, A. and {Saz Parkinson}, P.~M. and {Sgr{\`o}}, C. and {Siskind}, E.~J. and {Smith}, D.~A. and {Smith}, P.~D. and {Spinelli}, P. and {Starck}, J.-L. and {Stevenson}, M. and {Strickman}, M.~S. and {Suson}, D.~J. and {Takahashi}, H. and {Tanaka}, T. and {Thayer}, J.~B. and {Thompson}, D.~J. and {Tibaldo}, L. and {Tomsick}, J.~A. and {Torres}, D.~F. and {Tosti}, G. and {Tramacere}, A. and {Uchiyama}, Y. and {Usher}, T.~L. and {Vasileiou}, V. and {Vilchez}, N. and {Vitale}, V. and {Waite}, A.~P. and {Wang}, P. and {Wilms}, J. and {Winer}, B.~L. and {Wood}, K.~W. and {Ylinen}, T. and {Ziegler}, M.},
        title = "{Modulated High-Energy Gamma-Ray Emission from the Microquasar Cygnus X-3}",
      journal = {Science},
         year = 2009,
        month = dec,
       volume = {326},
       number = {5959},
        pages = {1512},
          doi = {10.1126/science.1182174},
       adsurl = {https://ui.adsabs.harvard.edu/abs/2009Sci...326.1512F}
}

@ARTICLE{2009Natur.462..620T,
       author = {{Tavani}, M. and {Bulgarelli}, A. and {Piano}, G. and {Sabatini}, S. and {Striani}, E. and {Evangelista}, Y. and {Trois}, A. and {Pooley}, G. and {Trushkin}, S. and {Nizhelskij}, N.~A. and {McCollough}, M. and {Koljonen}, K.~I.~I. and {Pucella}, G. and {Giuliani}, A. and {Chen}, A.~W. and {Costa}, E. and {Vittorini}, V. and {Trifoglio}, M. and {Gianotti}, F. and {Argan}, A. and {Barbiellini}, G. and {Caraveo}, P. and {Cattaneo}, P.~W. and {Cocco}, V. and {Contessi}, T. and {D'Ammando}, F. and {Del Monte}, E. and {de Paris}, G. and {Di Cocco}, G. and {di Persio}, G. and {Donnarumma}, I. and {Feroci}, M. and {Ferrari}, A. and {Fuschino}, F. and {Galli}, M. and {Labanti}, C. and {Lapshov}, I. and {Lazzarotto}, F. and {Lipari}, P. and {Longo}, F. and {Mattaini}, E. and {Marisaldi}, M. and {Mastropietro}, M. and {Mauri}, A. and {Mereghetti}, S. and {Morelli}, E. and {Morselli}, A. and {Pacciani}, L. and {Pellizzoni}, A. and {Perotti}, F. and {Picozza}, P. and {Pilia}, M. and {Prest}, M. and {Rapisarda}, M. and {Rappoldi}, A. and {Rossi}, E. and {Rubini}, A. and {Scalise}, E. and {Soffitta}, P. and {Vallazza}, E. and {Vercellone}, S. and {Zambra}, A. and {Zanello}, D. and {Pittori}, C. and {Verrecchia}, F. and {Giommi}, P. and {Colafrancesco}, S. and {Santolamazza}, P. and {Antonelli}, A. and {Salotti}, L.},
        title = "{Extreme particle acceleration in the microquasar CygnusX-3}",
      journal = {\nat},
         year = 2009,
        month = dec,
       volume = {462},
       number = {7273},
        pages = {620-623},
          doi = {10.1038/nature08578},
archivePrefix = {arXiv},
       eprint = {0910.5344},
 primaryClass = {astro-ph.HE},
       adsurl = {https://ui.adsabs.harvard.edu/abs/2009Natur.462..620T}
}

@ARTICLE{2024ApJDmytriiev,
       author = {{Dmytriiev}, Anton and {Zdziarski}, Andrzej A. and {Malyshev}, Denys and {Bosch-Ramon}, Valent{\'\i} and {Chernyakova}, Maria},
        title = "{Two Models for the Orbital Modulation of Gamma Rays in Cyg X-3}",
      journal = {\apj},
         year = 2024,
        month = sep,
       volume = {972},
       number = {1},
          eid = {85},
        pages = {85},
          doi = {10.3847/1538-4357/ad6440},
archivePrefix = {arXiv},
       eprint = {2405.09154},
 primaryClass = {astro-ph.HE},
       adsurl = {https://ui.adsabs.harvard.edu/abs/2024ApJ...972...85D}
}

@ARTICLE{2010MNRASDubus,
       author = {{Dubus}, G. and {Cerutti}, B. and {Henri}, G.},
        title = "{The relativistic jet of Cygnus X-3 in gamma-rays}",
      journal = {\mnras},
         year = 2010,
        month = may,
       volume = {404},
       number = {1},
        pages = {L55-L59},
          doi = {10.1111/j.1745-3933.2010.00834.x},
archivePrefix = {arXiv},
       eprint = {1002.3888},
 primaryClass = {astro-ph.HE},
       adsurl = {https://ui.adsabs.harvard.edu/abs/2010MNRAS.404L..55D}
}

@ARTICLE{2012MNRASZdziarski,
       author = {{Zdziarski}, Andrzej A. and {Sikora}, Marek and {Dubus}, Guillaume and {Yuan}, Feng and {Cerutti}, Benoit and {Ogorza{\l}ek}, Anna},
        title = "{The gamma-ray emitting region of the jet in Cyg X-3}",
      journal = {\mnras},
         year = 2012,
        month = apr,
       volume = {421},
       number = {4},
        pages = {2956-2968},
          doi = {10.1111/j.1365-2966.2012.20519.x},
archivePrefix = {arXiv},
       eprint = {1111.0878},
 primaryClass = {astro-ph.HE},
       adsurl = {https://ui.adsabs.harvard.edu/abs/2012MNRAS.421.2956Z}
}

@ARTICLE{2018MNRASZdziarski,
       author = {{Zdziarski}, Andrzej A. and {Malyshev}, Denys and {Dubus}, Guillaume and {Pooley}, Guy G. and {Johnson}, Tyrel and {Frankowski}, Adam and {De Marco}, Barbara and {Chernyakova}, Maria and {Rao}, A.~R.},
        title = "{A comprehensive study of high-energy gamma-ray and radio emission from Cyg X-3}",
      journal = {\mnras},
         year = 2018,
        month = oct,
       volume = {479},
       number = {4},
        pages = {4399-4415},
          doi = {10.1093/mnras/sty1618},
archivePrefix = {arXiv},
       eprint = {1804.07460},
 primaryClass = {astro-ph.HE},
       adsurl = {https://ui.adsabs.harvard.edu/abs/2018MNRAS.479.4399Z}
}

@ARTICLE{1996AJ....112.2690W,
       author = {{Waltman}, E.~B. and {Foster}, R.~S. and {Pooley}, G.~G. and {Fender}, R.~P. and {Ghigo}, F.~D.},
        title = "{Quenched Radio Emission in Cygnus X-3}",
      journal = {\aj},
         year = 1996,
        month = dec,
       volume = {112},
        pages = {2690},
          doi = {10.1086/118213},
       adsurl = {https://ui.adsabs.harvard.edu/abs/1996AJ....112.2690W}
}

@ARTICLE{2006MNRAS.367.1432M,
       author = {{Miller-Jones}, J.~C.~A. and {Fender}, R.~P. and {Nakar}, E.},
        title = "{Opening angles, Lorentz factors and confinement of X-ray binary jets}",
      journal = {\mnras},
         year = 2006,
        month = apr,
       volume = {367},
       number = {4},
        pages = {1432-1440},
          doi = {10.1111/j.1365-2966.2006.10092.x},
archivePrefix = {arXiv},
       eprint = {astro-ph/0601482},
 primaryClass = {astro-ph},
       adsurl = {https://ui.adsabs.harvard.edu/abs/2006MNRAS.367.1432M}
}

@ARTICLE{2007ARA&A..45..177C,
       author = {{Crowther}, Paul A.},
        title = "{Physical Properties of Wolf-Rayet Stars}",
      journal = {\araa},
         year = 2007,
        month = sep,
       volume = {45},
       number = {1},
        pages = {177-219},
          doi = {10.1146/annurev.astro.45.051806.110615},
archivePrefix = {arXiv},
       eprint = {astro-ph/0610356},
 primaryClass = {astro-ph},
       adsurl = {https://ui.adsabs.harvard.edu/abs/2007ARA&A..45..177C}
}

@ARTICLE{2011A&A...529A.120C,
       author = {{Cerutti}, B. and {Dubus}, G. and {Malzac}, J. and {Szostek}, A. and {Belmont}, R. and {Zdziarski}, A.~A. and {Henri}, G.},
        title = "{Absorption of high-energy gamma rays in Cygnus X-3}",
      journal = {\aap},
         year = 2011,
        month = may,
       volume = {529},
          eid = {A120},
        pages = {A120},
          doi = {10.1051/0004-6361/201116581},
archivePrefix = {arXiv},
       eprint = {1103.3875},
 primaryClass = {astro-ph.HE},
       adsurl = {https://ui.adsabs.harvard.edu/abs/2011A&A...529A.120C}
}

@article{1959On,
  author  = {Kulikov, G. V. and Khristiansen, G. B.},
  title   = {On the Size Spectrum of Extensive Air Showers},
  journal = {Soviet Physics JETP},
  volume  = {35},
  pages   = {441},
  year    = {1959}
}

@ARTICLE{2019ApJ...871..244A,
       author = {{Antokhin}, Igor I. and {Cherepashchuk}, Anatol M.},
        title = "{The Period Change of Cyg X-3}",
      journal = {\apj},
         year = 2019,
        month = feb,
       volume = {871},
       number = {2},
          eid = {244},
        pages = {244},
          doi = {10.3847/1538-4357/aafb38},
archivePrefix = {arXiv},
       eprint = {1807.00817},
 primaryClass = {astro-ph.HE},
       adsurl = {https://ui.adsabs.harvard.edu/abs/2019ApJ...871..244A}
}

@ARTICLE{2018MNRAS.481.1455K,
       author = {{Khangulyan}, Dmitry and {Bosch-Ramon}, Valent{\'\i} and {Uchiyama}, Yasunobu},
        title = "{Inverse Compton emission from relativistic jets in binary systems}",
      journal = {\mnras},
         year = 2018,
        month = dec,
       volume = {481},
       number = {2},
        pages = {1455-1468},
          doi = {10.1093/mnras/sty2356},
archivePrefix = {arXiv},
       eprint = {1808.09628},
 primaryClass = {astro-ph.HE},
       adsurl = {https://ui.adsabs.harvard.edu/abs/2018MNRAS.481.1455K}
}

@ARTICLE{2014ApJ...783..100K,
       author = {{Khangulyan}, D. and {Aharonian}, F.~A. and {Kelner}, S.~R.},
        title = "{Simple Analytical Approximations for Treatment of Inverse Compton Scattering of Relativistic Electrons in the Blackbody Radiation Field}",
      journal = {\apj},
         year = 2014,
        month = mar,
       volume = {783},
       number = {2},
          eid = {100},
        pages = {100},
          doi = {10.1088/0004-637X/783/2/100},
archivePrefix = {arXiv},
       eprint = {1310.7971},
 primaryClass = {astro-ph.HE},
       adsurl = {https://ui.adsabs.harvard.edu/abs/2014ApJ...783..100K}
}
\bibliographystyle{aasjournal}

\end{document}